\documentclass[iop,apjl,twocolappendix,numberedappendix,appendixfloats]{emulateapj}

\usepackage{amsmath,amssymb}
\usepackage{amsmath}
\usepackage{bm}
\usepackage{color}

\shorttitle{Extremely red  SMGs}
\shortauthors{Ikarashi et al.}

\begin{document}

\title{Extremely Red Submillimeter Galaxies:  New $z\gtrsim4$--6 Candidates Discovered using ALMA and Jansky VLA}
\author{Soh~Ikarashi\altaffilmark{1}, R.\,J.~Ivison\altaffilmark{2,3}, Karina\,I.~Caputi\altaffilmark{1}, Koichiro~Nakanishi\altaffilmark{4,5,6}, Claudia\,D.~P.~Lagos\altaffilmark{7}, M.\,L\,.\,N.~Ashby\altaffilmark{8}, Itziar~Aretxaga\altaffilmark{9}, James\,S.~Dunlop\altaffilmark{2}, Bunyo~Hatsukade\altaffilmark{4}, David\,H.~Hughes\altaffilmark{9}, Daisuke~Iono\altaffilmark{4,5}, Takuma~Izumi\altaffilmark{10}, Ryohei~Kawabe\altaffilmark{4,5}, Kotaro~Kohno\altaffilmark{10,11}, Kentaro~Motohara\altaffilmark{10}, Kouji~Ohta\altaffilmark{12}, Yoichi~Tamura\altaffilmark{10}, Hideki~Umehata\altaffilmark{10,13}, Grant\,W.~Wilson\altaffilmark{14}, Kiyoto~Yabe\altaffilmark{15}, Min\,S.~Yun\altaffilmark{14} \\
{\scriptsize 
$^1${Kapteyn Astronomical Institute, University of Groningen, P.\,O.~Box 800, 9700 AV Groningen, The Netherlands} \\
$^2${Institute for Astronomy, University of Edinburgh, Royal Observatory, Blackford Hill, Edinburgh EH9 3HJ, UK} \\
$^3${European Southern Observatory, Karl Schwarzschild Str.~2, D-85748 Garching, Germany}  \\
$^4${National Astronomical Observatory of Japan, Mitaka, Tokyo 181-8588, Japan} \\
$^5${The Graduate University for Advanced Studies (SOKENDAI), 2-21-1 Osawa, Mitaka, Tokyo 181-8588, Japan}  \\
$^6${Joint ALMA Observatory, Alonso de Cordova 3107, Vitacura, Santiago 763 0355, Chile} \\
$^7${International Centre for Radio Astronomy Research, University of Western Australia, 7 Fairway, Crawley 6009, Perth WA, Australia} \\
$^8${Harvard-Smithsonian Center for Astrophysics, 60 Garden St., Cambridge, MA 02138, USA} \\
$^9${Instituto Nacional de Astrof\'{\i}sica, \'Optica y Electr\'onica (INAOE), Aptdo.\ Postal 51 y 216, 72000 Puebla, Mexico} \\
$^{10}${Institute of Astronomy, University of Tokyo, 2-21-1 Osawa, Mitaka, Tokyo 181-0015, Japan}  \\
$^{11}${Research Center for the Early Universe, School of Science, University of Tokyo, 7-3-1 Hongo, Bunkyo, Tokyo 113-0033, Japan} \\
$^{12}${Department of Astronomy, Kyoto University, Kitashirakawa-Oiwake-Cho, Sakyo-ku, Kyoto 606-8502, Japan} \\
$^{13}${The Open University of Japan, 2-11 Wakaba, Mihama-ku, Chiba 261-8586} \\
$^{14}${Department of Astronomy, University of Massachusetts, Amherst, MA 01003, USA} \\
$^{15}${Kavli Institute for the Physics and Mathematics of the Universe (WPI), The University of Tokyo, 5-1-5 Kashiwanoha, Kashiwa, Chiba, 277-8583, Japan}} }

\begin{abstract}
  We present the detailed characterization of two extremely red
  submillimeter galaxies (SMGs), ASXDF1100.053.1 and 231.1, with the
  Atacama Large Millimeter/submillimeter Array (ALMA) and the Jansky
  Very Large Array (VLA). These SMGs were selected originally using
  AzTEC at 1100\,$\mu$m, and are observed by {\it Herschel} to be
  faint at 100--500\,$\mu$m. Their (sub)millimeter colors are as red
  as -- or redder -- than known $z\gtrsim5$ SMGs; indeed,
  ASXDF1100.053.1 is redder than HFLS\,3, which lies at $z=6.3$. They
  are also faint and red in the near-/mid-infrared: $\sim1\,\mu$Jy at
  IRAC 4.5\,$\mu$m and $<0.2\,\mu$Jy in the $K_{\rm s}$ filter. These
  SMGs are also faint in the radio waveband, where
  $F_{\rm 6GHz}=4.5\,\mu$Jy for ASXDF1100.053.1 and
  $F_{\rm 1.4GHz}=28\,\mu$Jy for ASXDF1100.231.1, suggestive of
  $z=6.5^{+1.4}_{-1.1}$ and $z=4.1^{+0.6}_{-0.7}$ for ASXDF1100.053.1
  and 231.1, respectively. ASXDF1100.231.1 has a flux excess in the
  3.6-$\mu$m filter, probably due to H$\alpha$ emission at
  $z=4$--5. Derived properties of ASXDF1100.053.1 for
    $z=5.5$--7.5 and 231.1 for $z=3.5$--5.5 are as follows: their
  infrared luminosities are $[6.5-7.4]\times10^{12}$ and
  $[4.2-4.5]\times10^{12} {\rm L_{\odot}}$; their stellar masses are
  $[0.9-2]\times10^{11}$ and $[0.4-3]\times10^{10} {\rm M_{\odot}}$;
  their circularized half-light radii in the ALMA maps are $\sim1$ and
  $\lesssim0.2$\,kpc ($\sim$2--3\,kpc for 90\% of the total
  flux). Lastly, their surface infrared luminosity densities,
  $\Sigma_{\rm IR}$, are  $\sim1$$\times10^{12}$ and
  $\gtrsim1.5$$\times10^{13}$ ${\rm L_{\odot}}$\,kpc$^{-2}$, similar
  to values seen for local (U)LIRGs. These data suggest that
  ASXDF1100.053.1 and 231.1 are compact SMGs at $z\gtrsim4$ and can
  plausibly evolve into $z\gtrsim3$ compact quiescent galaxies.
\end{abstract}


\section{Introduction}

Submillimeter (submm) galaxies (SMGs) with infrared (IR, rest-frame
8--1000\,$\mu$m) luminosities,
$L_{\rm IR}\gtrsim 10^{12}$\,L$_{\odot}$, are routinely detected in
deep continuum images at $\lambda_{\rm obs}=850$--1300\,$\mu$m using
ground-based single-dish telescopes. Even out to $z\sim7$, there is no
significant loss of sensitivity to these SMGs, given the strong
negative $K$-correction in the Rayleigh-Jeans tail of their dust
spectral energy distributions (SEDs) \citep[e.g.][]{bla02}.

Despite 20 years of deep submm surveys since \citet{sib97},
our knowledge of the upper half of the redshift distribution of SMGs
remains incomplete. Early attempts to determine redshifts were
conducted towards SMGs with radio counterparts, because low-resolution
(sub)mm images obtained with single dishes require
high-resolution radio continuum maps from radio interferometers such
as the Jansky Very Large Array (VLA) in order to pinpoint source
positions \citep{ivi98, ivi00, ivi02, ivi05, ivi07, sma99, bor04,
  pop06, are11, big11, yun12}. Intensive studies of radio-bright SMGs
were able to yield spectroscopic redshifts for those out to $z\sim3$
\citep[e.g.][]{chap03, cha05}. However, at that time radio
sensitivities could not detect SMGs beyond $z\sim3$, and as many as
half of SMGs lacked reliable radio counterparts (see e.g.\
\citealt{ivi07,big11}, cf. \citealt{lin11}). Later attempts to
determine SMG positions and redshifts using near- and mid-IR imaging
could not fully overcome the bias towards lower redshifts, since the
$K$ corrections there are no more favorable than those in the radio
regime, such that high-redshift sources are much fainter
\citep[e.g.][]{war11, yun12}. Millimeter spectroscopic
surveys toward gravitationally-lensed, dusty, star-forming galaxies,
taking advantage of their apparent ultra brightness, revealed a
redshift distribution stretching out to $z\sim5.8$
\citep[e.g.][]{vie13,wei13,str16}. These surveys suggested a larger
fraction of SMGs at $z\gtrsim3$ than previous studies of unlensed
SMGs, perhaps partly because they were selected at 1.3\,mm rather than
the traditional 0.8--1.1\,mm, but also because the requirement for
high magnification favors galaxies with a long line of sight. We need
to reveal the intrinsic redshift distributions of unlensed SMGs in
large contiguous maps to determine their abundance in the early
Universe and to study the evolution of the most massive galaxies via
abundance matching with other populations, and with cosmological
predictions \citep[e.g.][]{hay13,wil15}. 

Early (sub)mm interferometric imaging of intrinsically bright
SMGs, conducted with the IRAM Plateau de Bure interferometer (PdBI)
and the Submillimeter Array (SMA) \citep[e.g.][]{gea00, ion06, you07,
  dan08, you09}, pinpointed the positions of SMGs, including
radio-faint ones, and resulted in the discovery of SMGs at
$z\gtrsim4$--5 \citep[e.g.][]{capa11}.  Subsequently, surveys with
PdBI and the Combined Array for Research in Millimeter-wave Astronomy
(CARMA) indicated that the redshift distribution of intrinsically
bright SMGs most likely stretches to $z\sim6$ \citep{smo12}.

The capabilities of the Atacama Large Millimeter/submm
Array (ALMA) now enable astronomers to rapidly pinpoint the
positions of large samples of SMGs, with no strong biases
\citep[though see][]{zha16}. ALMA submm continuum imaging
surveys towards LABOCA 870-$\mu$m-selected SMGs
\citep[e.g.][]{hod13,sim14} and AzTEC 1100-$\mu$m-selected SMGs
\citep{ika14} have uncovered a number of radio-faint SMGs. Some
of these radio-faint SMGs have been too faint at optical/near-IR
wavelengths to permit estimation of their redshifts using
standard techniques \citep{sim14,ika14}. Some could lie at very
high redshifts, i.e.\ $z\gtrsim5$; alternatively, they could be
heavily dust-obscured SMGs at more moderate redshifts,
$z\approx 3$--5. The redshifts of these SMGs remains a puzzle,
with important implications for our understanding of early galaxy
evolution.

ALMA mm-wave continuum imaging of $z\gtrsim3$ candidate SMGs
have revealed surprisingly compact sizes, supporting the idea
that $z\gtrsim3$ SMGs could evolve into compact quiescent
galaxies at $z\sim2$ \citep{ika14}. The latest intensive
optical/near-/mid-IR extragalactic surveys have reported compact
quiescent galaxies up to $z\sim4$ \citep{str15}. In order to
understand the formation phase of these massive, passive galaxies
at $z\gtrsim3$, surveys and studies of SMGs $z\gtrsim4$--5 are as
important today as they ever were.

In this paper, we present a detailed multi-wavelength analysis of
two ALMA-identified galaxies, ASXDF1100.053.1 and
ASXDF1100.231.1, detected originally in a deep ASTE/AzTEC survey
at 1100\,$\mu$m \citep{ika14}. These SMGs were selected for
further scrutiny on the basis of their secure non-detections in
{\it Herschel} 100--500-$\mu$m images, which give the most useful
constraints on redness at submm wavelengths \citep[see
also][]{cox11,rie13,dow14,asb16,ivi16}. Too faint at
optical/near-IR wavelengths to allow meaningful estimation of
their redshifts using classical photometric techniques, we have
instead determined photometric redshifts using deep
radio/submm/far-IR images from the Janksy VLA, ALMA,
SCUBA-2 and {\it Herschel}, respectively, aiming to reveal
whether these galaxies are indeed located at very high redshifts
--- obvious candidate progenitors of the massive passive galaxies
at $z\gtrsim3$. We adopt throughout a cosmology with
$H_{\rm 0}= 70$\,km\,s$^{-1}$\,Mpc$^{-1}$, $\Omega_{\rm M}= 0.3$
and $\Omega_{\rm \Lambda}= 0.7$, and all magnitudes refer to the
AB system.

\section{The targets: ASXDF1100.053.1 and 231.1}
\label{sec1}

\begin{figure}
\epsscale{1.0}
\plotone{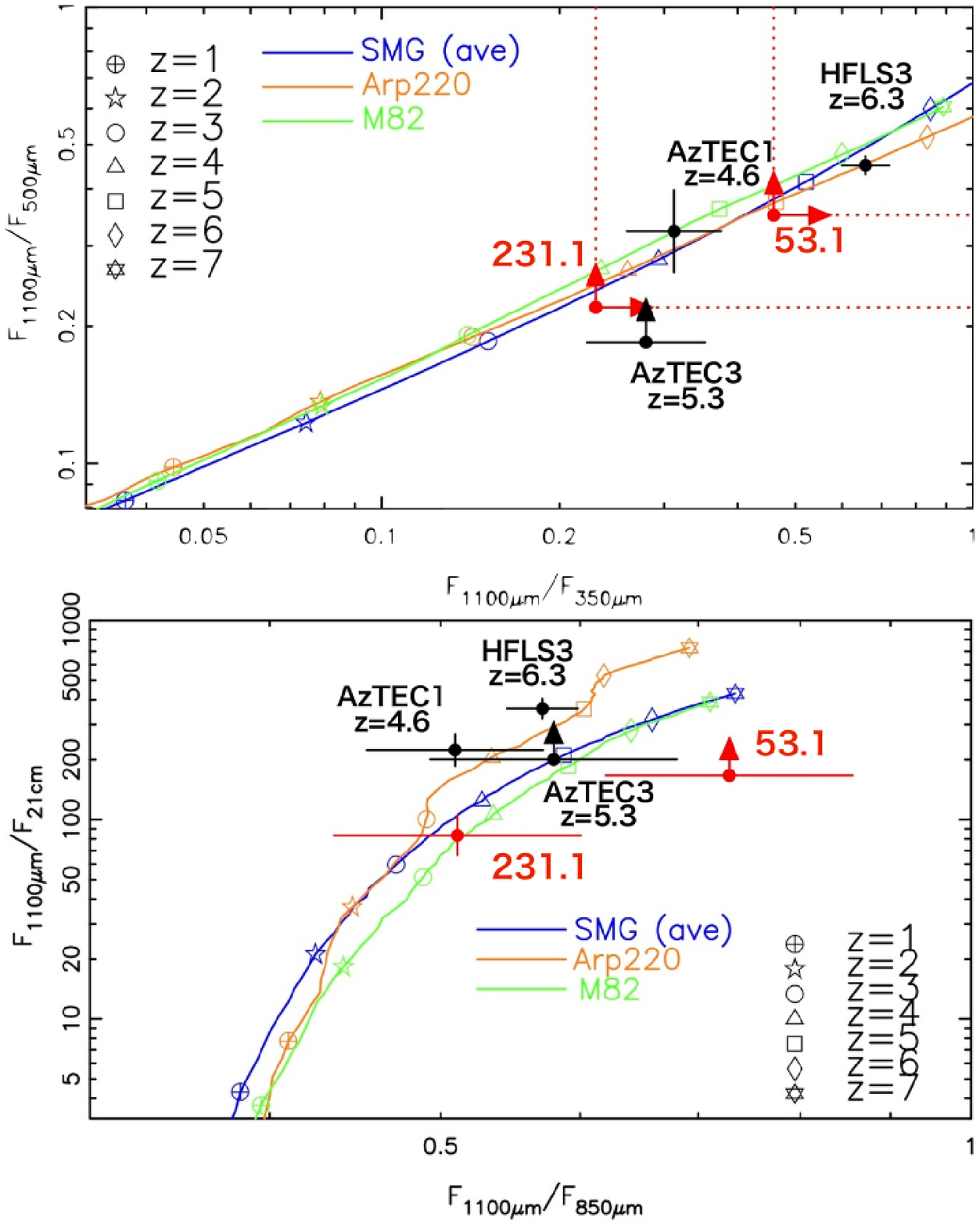}
\caption{Comparison of submm/mm/radio colors for
  ASXDF1100.053.1 and 231.1 with colors of known $z\gtrsim5$ SMGs from
  the literature.  Submm/mm/radio colors for
  ASXDF1100.053.1 and 231.1 are marked by red crosses or arrows based
  on fluxes in Table.~\ref{tbl-1}.  The highest-redshift SMG known,
  HFLS3 at $z=6.3$ \citep{rie13}, and known $z\sim5$ SMGs, AzTEC1 and
  AzTEC 3 \citep{smo15}, are marked by black points.  The blue, orange
  and green lines mark the color track as a function of redshift of
  the average SED of 99 ALMA-identified SMGs \citep{swi14}, SED templates
  of Arp\,220 and M\,82 \citep{sil98}, respectively.  }
\label{fig:submmcol}
\end{figure}

ASXDF1100.053.1 and 231.1 are the brightest and second-brightest
1100-$\mu$m-selected ALMA-identified SMGs among the $z\gtrsim3$
candidates discovered in our ALMA Cycle-1 program
(2012.1.00326.S: PI.\ Ikarashi). The parent sample consists of
221 SMGs discovered in a deep AzTEC/ASTE 1100-$\mu$m map covering
950\,arcmin$^2$ of the Subaru/{\it XMM-Newton} Deep Field (SXDF)
\citep[e.g.][]{fur08}, which includes the UKIDSS Ultra Deep
Survey (UDS) field \citep[e.g.][]{law07}. In our ALMA program, we
targeted 30 SMGs from this parent sample, selected on the basis
of their faintness in 1.4-GHz VLA imaging
($5\sigma\lesssim 35\,\mu$Jy, \citealt{aru16}) and SPIRE
250-$\mu$m images ($3\sigma_{\rm confusion}\lesssim 18.3$\,mJy,
\citealt{oli12}), aiming to reveal the tail of the SMG redshift
distribution. The faintness of these two SMGs at
optical/near-/mid-IR wavelengths suggests $z\gtrsim4$--5
(Ikarashi et al.\ 2015; Fig.~1). The submm
(250, 350, 500 and 850\,$\mu$m)/mm (1100\,$\mu$m)/radio
(1.4\,GHz) colors of ASXDF1100.053.1 and 231.1 are as red as --
or redder -- than known $z\gtrsim5$ SMGs, which suggests that
these new SMGs could lie at $z\gtrsim5$
(Fig.~\ref{fig:submmcol}). We thus focus on these two SMGs for a
pilot study of candidate extremely high-redshift SMGs. 

\section{Data and photometry}

Here we describe the observational data used in this paper. Our
images are shown in Figs~\ref{fig:stamp53} and
\ref{fig:stamp231}, and measurements are listed in
Table.~\ref{tbl-1}.

\subsection{ALMA 1100-$\mu$m continuum}

We first describe the ALMA data taken in Cycle~1
\citep[S.~Ikarashi\,et\,al.\, in preparation: see also][]{ika14}.
These observations were carried out with an array configuration
similar to C32-3, with 25 working 12-m antennas covering $uv$
distances up to $\sim 400$\,k$\lambda$. On-source observation
times (per target) were 3.6--4.5 minutes.
 
The two SMGs were also observed as part of an ALMA continuum
imaging survey of 333 bright AzTEC SMGs in Cycle~2 (2013.1.00781:
PI.\ Hatsukade). These observations were carried out in array
configurations C34-5 and C34-7, with 37--38 working 12-m antennas
covering $uv$ distances up to $\sim 1500$\,k$\lambda$. On-source
observation times per source were 0.6 minutes.

We combined the ALMA data obtained in Cycles 1 and 2. Synthesized
beams were then 0$''$.46$\times$0$''$.35 (PA, 69$^{\circ}$) and
0$''$.57$\times$0$''$.48 (PA, 82$^{\circ}$) for ASXDF1100.053.1
and 231.1, respectively, with sensitivities of 70 and
63\,$\mu$Jy\,beam$^{-1}$ (1$\sigma$). ASXDF1100.053.1 and 231.1
were detected with $S_{\rm peak}/N=$ 27 and 29, respectively,
with total flux densities, $F_{\rm 1100 \mu m}=3.51\pm0.15$ and
$2.28\pm0.08$\,mJy.

Both ASXDF1100.053.1 and 231.1 appear to be single, unblended
SMGs, with no signs of multiplicity; their ALMA 1100-$\mu$m flux
densities are consistent (within 1$\sigma$) with those measured
by AzTEC/ASTE (S.\,Ikarashi\,et\,al.~in preparation).

\subsection{Jansky Very Large Array radio continuum}

\subsubsection{Classic VLA 1.4-GHz continuum}

The accurate SMG positions from our ALMA images enable us to
exploit existing deep VLA radio continuum maps. ASXDF1100.231.1
was detected at 3.3$\sigma$ in an existing wide, deep VLA 1.4-GHz
image of the SXDF field \citep{aru16}; ASXDF1100.053.1 was not
detected. The r.m.s.\ noise of the map is
6--8\,$\mu$Jy\,beam$^{-1}$, and the FWHM synthesized beam is
$\sim1''$.5.

\subsubsection{Jansky VLA 6-GHz continuum}

In order to measure the radio flux density of ASXDF1100.053.1, we
conducted new extremely deep Jansky VLA observations. The data
were obtained from 2015 February to April with the Jansky VLA in
its B configuration, using the new 3-bit samplers\footnote{We
  acknowledge funding towards the 3-bit samplers used in this
  work from ERC Advanced Grant 321302, COSMICISM.}, with the
WIDAR correlator, covering an almost contiguous 4-GHz band across
4--8\,GHz (several spectral windows covering a total of
$\approx 0.25$\,GHz were discarded due to radio-frequency
interference). The phase center was set to be the position of
ASXDF1100.053.1. The FWHM field of view (FoV) covers a circular
area of radius 3.7\,arcmin in the final map. The total
observation time was 14\,hr, of which 10.1\,hr were spent
on-source. We chose J0239$-$0234 as the gain calibrator, using
3C\,48 as the bandpass calibrator and to set the flux density
scale. We reduced the data with CASA and imaged using a natural
weighting scheme. The resulting map reaches an r.m.s.\ noise
level of 1.1\,$\mu$Jy\,beam$^{-1}$ and has a synthesized beam
size of $1''.5\,\times1''.2$ (PA, $16^{\circ}.2$). Given the
color correction between $\nu_{\rm obs}=$1.4 and 6\,GHz for a
radio spectral index, $\alpha=-0.8$, the sensitivity of the new
Jansky VLA 6-GHz map is more than $2\times$ deeper than the old
VLA 1.4-GHz map. In the new 6-GHz map we detect emission at the
position of ASXDF1100.053.1: $4.5\pm1.1\,\mu$Jy (4.0\,$\sigma$).
The source characteristics are summarized in
Table~\ref{tbl:jvla}.

\subsection{{\it Herschel}/SPIRE 250--500-$\mu$m continuum}

We use the {\it Herschel}/SPIRE 250, 350 and 500-$\mu$m maps in
the UKIDSS UDS field, provided as part of the HerMES
\citep{oli12} 2nd data release (DR2). Armed with ALMA positions,
accurate to $<0.1''$, it is clear that both of ASXDF1100.053.1
and 231.1 were not detected in deep imaging by {\it Herschel}
PACS and SPIRE images (see Figs~\ref{fig:stamp53} and
\ref{fig:stamp231}): the respective flux densities in the 250-,
350- and 500-$\mu$m maps are 4.2, 6.3 and 9.5\,mJy\,beam$^{-1}$,
and those at the position of ASXDF1100.231.1 are 0.7, 5.2 and
3.1\,mJy\,beam$^{-1}$. These values are below the 3-$\sigma$
limits measured in residual SPIRE maps of $5'\times5'$ areas
around ASXDF1100.053.1 and 231.1, where the residual maps have
been deblending based on the positions of known VLA-1.4 GHz and
MIPS 24-$\mu$m sources. The respective 250-, 350- and 500-$\mu$m
flux densities of ASXDF1100.053.1 in the residual images are
$-1.3$, 0.1 and 4.3\,mJy\,beam$^{-1}$, and those of
ASXDF1100.231.1 are 0.6, 4.0 and 0.8 mJy\,beam$^{-1}$.

\subsection{SCUBA-2 850-$\mu$m continuum}

Both ASXDF1100.053.1 and 231.1 are detected in the deep SCUBA-2
850-$\mu$m map of the SCUBA-2 Cosmology Legacy Survey Data
Release 1. In \citet{gea16} they are referred to as UDS0186 and
UDS0206, with 850-$\mu$m flux densities of $4.8\pm1.1$ and
$4.5\pm1.1$ mJy, respectively. The respective offsets between
their ALMA 1100-$\mu$m and SCUBA-2 850-$\mu$m positions are 2.5
and 5.7 arcsec, consistent with the SCUBA-2 positional offset
distributions \citep{sim15b}. 

\subsection{{\it Spitzer} mid-IR continuum}

We use the deep {\it Spitzer} IRAC 3.6- and 4.5-$\mu$m maps from
the {\it Spitzer} Extended Deep Survey \citep[SEDS; ][]{ash13}
and IRAC 5.8- and 8.0-$\mu$m and MIPS 24-$\mu$m data from the
{\it Spitzer} UKIDSS Ultra Deep Survey \citep[SpUDS; PI.\
J.~Dunlop; see e.g.][]{cap11}. IRAC counterparts of
ASXDF1100.053.1 and 231.1 were found at (RA, Dec) $=$
(02$^{\rm h}$16$^{\rm m}$48.19$^{\rm s}$,
$-04^{\rm \circ}$58$'$59.6$''$) and (02$^{\rm
  h}$17$^{\rm m}$59.62$^{\rm s}$,
$-04^{\rm \circ}$46$'$59.7$''$), respectively, with offsets from
the ALMA positions of 0$''$.2 and 0$''$.5. Photometric
measurements performed at the IRAC positions with fixed apertures
and aperture corrections for 2.4$''$-$\phi$(IRAC 3.6 and
4.5\,$\mu$m); 2.8$''$-$\phi$ (IRAC 5.8 and 8.0\,$\mu$m). The MIPS
24-$\mu$m upper limit ($3\sigma$) is based on photometry at random positions
including an aperture correction for
7$''$-$\phi$. For $<$2-$\sigma$ detections in IRAC maps, we adopt
$2\sigma$ upper limits. 

\subsection{Optical/near-IR continuum} 

We use optical/near-IR images at $B$, $V$, $Rc$, $i'$ and
$z'$-bands from the Subaru Telescope \citep{fur08} and near-IR
images at $J$, $H$ and $K_{\rm s}$-bands from the UKIRT IR
Deep Sky Survey \citep[UKIDSS;][]{law07}. We measured fluxes with
fixed apertures at the positions of the IRAC counterparts and
applied aperture corrections: 2$''$-$\phi$ aperture for $B$
through $K_{\rm s}$. Errors were derived from random aperture
photometry. Again, For $<$2-$\sigma$ detections, we adopt
$2\sigma$ upper limits.

\begin{figure*}
  \epsscale{1.} \plotone{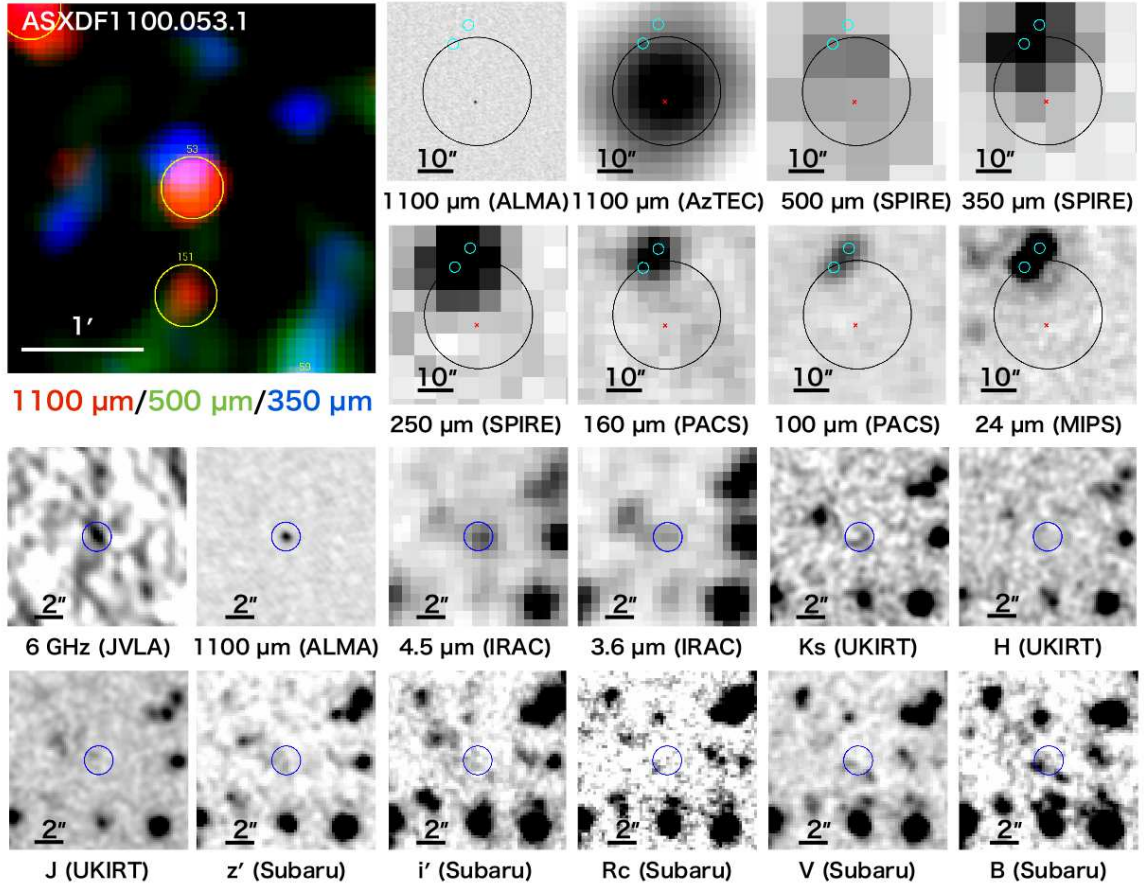}
  \caption{Multi-wavelength images of ASXDF1100.053.1.  {\it Top
      right:} RGB image (R, G and B being 1100, 500 and 350\,$\mu$m,
    respectively) around ASXDF1100.053.1.  Yellow circles mark AzTEC
    1100-$\mu$m sources.  {\it First and second rows from top:} ALMA,
    AzTEC, SPIRE, PACS and {\it Spitzer} images. The black circle
    marks the AzTEC position of ASXDF1100.053.1 and the beam size of
    the AzTEC/ASTE image (30$''$).  The red cross marks the ALMA
    position of ASXDF1100.053.1.  The small cyan circle marks the
    position of a {\it Spitzer} and {\it Herschel} bright source near
    ASXDF1100.053.1.  {\it Third and fourth rows from top:} Jansky
    VLA, ALMA, IRAC, UKIRT and Subaru images of ASXDF1100.053.1. The
    blue circle marks the ALMA position of
    ASXDF1100.053.1.} \label{fig:stamp53} \end{figure*}

\begin{figure*}
\epsscale{1.}
\plotone{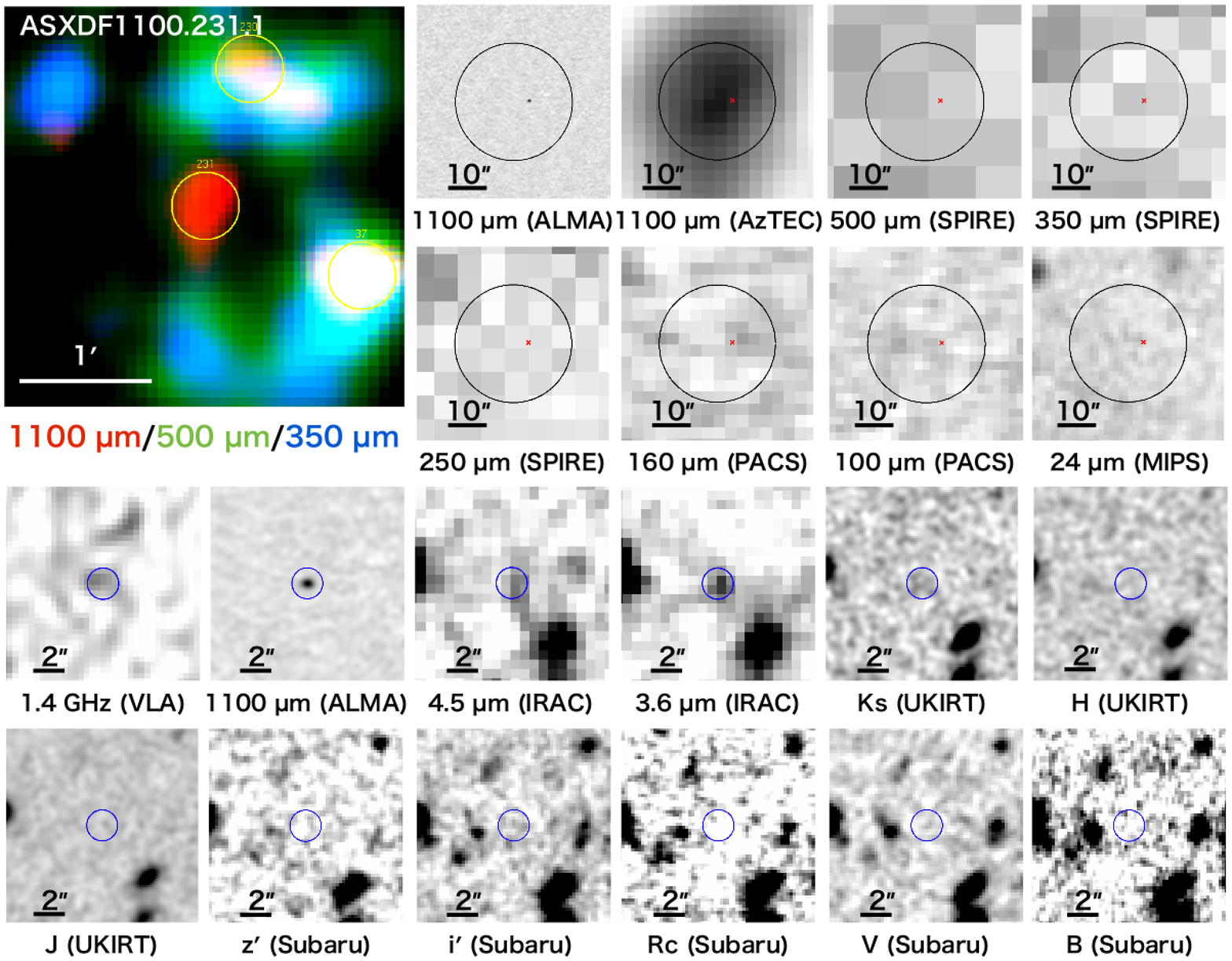}
\caption{Multi-wavelength images of ASXDF1100.231.1.} \label{fig:stamp231}
\end{figure*}

\section{Radio/mm-wave photometric redshifts} 
 
The existence of SMGs which are extremely faint at optical to
mid-IR wavelengths has long been recognized
\citep[e.g.][]{hug98,ivi00,wan09, wei09a, wal12} and
radio/submm colors have been used to estimate the
redshifts of heavily dust-obscured SMGs \citep[e.g.][]{car99,
  hug02, ivi05, are03, are05, are07}, exploiting the tight
correlation between radio and far-IR luminosities seen for local
galaxies \citep{con92}.

\begin{table*}
\begin{center}
\caption{Photometric Data of ASXDF1100.053.1 and 231.1.\label{tbl-1}}
\scalebox{1.0}{ 
\begin{tabular}{c c c c}
\tableline \tableline 
                                                                             &    ASXDF1100.053.1                                  &      ASXDF1100.231.1              &                   \\
Wavelength                                                          &      Flux ($\mu$Jy)                                     &    Flux ($\mu$Jy)          & Reference  \\ \tableline
   Suprime Cam $B$-band (0.45 $\mu$m)          &            $<$  0.014        				   &   $<$  0.016                &       1            \\      
   Suprime Cam $V$-band (0.55 $\mu$m)          &           $<$  0.023          				   &  $<$ 0.021                 &       1            \\    
   Suprime Cam $Rc$-band (0.66 $\mu$m)        &          $<$  0.027                			   & $<$ 0.025                  &       1            \\      
   Suprime Cam $i'$-band (0.77 $\mu$m)           &            $<$ 0.027          		   		   &  $<$ 0.025          &       1            \\     
   Suprime Cam $z'$-band (0.92 $\mu$m)          &            $0.067\pm0.036$                           &  $<$ 0.064           &       1            \\  
   WFCAM $J$-band (1.2 $\mu$m)                     &             $<$0.14                    			   &   $<$0.14   &       2            \\ 
   WFCAM $H$-band (1.6 $\mu$m)                   &             $<$0.23                       		   & $<$0.23   &       2            \\ 
   WFCAM $K_{\rm s}$ (2.2 $\mu$m)                   &             $<$0.19                       		  &  $<$0.19   &       2            \\ 
   IRAC     3.6  $\mu$m                                       &           $0.61\pm0.14$                		   &   $1.00\pm0.17$     &       3            \\  
   IRAC     4.5  $\mu$m                                       &          $1.43\pm0.17$                  		  &    $0.93\pm0.22$         &       3            \\  
   IRAC     5.8  $\mu$m                                       &              $3.5\pm1.9$                        	  &    $<$ 3.5     &  4            \\  
   IRAC     8.0  $\mu$m                                       &            $<$ 4.7                                         &   $7.4\pm2.6$   &       4            \\  
   MIPS     24  $\mu$m                                        &           $<$ 66                   			  &       $<$ 66                     &       4            \\  
   PACS   110  $\mu$m                                        &              $<$ 2400             			 &     $<$ 2400                &       5            \\  
   PACS   160  $\mu$m                                       &              $<$ 5000              			  &    $<$ 5000               &       5            \\  
   SPIRE  250  $\mu$m                                       &               $<$ 9600               			 &          $<$ 8800       &       5            \\  
   SPIRE  350  $\mu$m                                       &              $<$  7700               			  &  $<$  9800            &       5            \\  
   SPIRE  500  $\mu$m                                       &               $<$ 10000                 			  &    $<$ 10000         &       5            \\ 
  SCUBA2  850  $\mu$m                                       &         $4800 \pm 1100$  		  &      $4500\pm1100$       &       6          \\ 
   ALMA  1100  $\mu$m                                       &    $3510 \pm 150$                  		  &   $2280\pm 80$        &       7            \\     
   JVLA 6 GHz                                                        &        $4.46\pm1.1$                 		&    \nodata    &       7            \\        
   VLA  1.4 GHz                                                    &            $<$17.8                    			&      $27.6\pm8.7$                 &    8            \\            
\tableline 
\multicolumn{4}{l}{{\bf Notes.} 2- and 3-$\sigma$ upper limits are presented for (stellar) emission at 0.45--8.0\,$\mu$m and dust/synchrotron } \\
\multicolumn{4}{l}{emission at 24\,$\mu$m through 1.4\,GHz, respectively.}  \\
\multicolumn{4}{l}{References: (1) \citet{fur08}; (2)  \citet{law07}; (3) \citet{ash13}; (4) \citet{cap11},} \\
\multicolumn{4}{l}{(5) \citet{oli12}; (6) \citet{gea16}; (7) this work; (8) \citet{aru16}. } \\
\end{tabular} }
 \end{center}
\end{table*}

\begin{table}
\begin{center}
\caption{JVLA observations.\label{tbl:jvla}}
\begin{tabular}{c c}
\tableline \tableline 
  Observation date                                                               &    2015 February 16             \\
                                                                                            &  March 2, 9, 17 and 30    \\
                                                                                            &    April 2        \\
  Frequency                                                        &      4--8\,GHz                   \\  
  Phase center (J2000)                                                         &     RA\,$=$\,02$^{\rm h}$16$^{\rm m}$48$^{\rm s}$                   \\
                                                                                              &  Dec.\,$=$\,$-04^{\rm \circ}$58$'$59$''$                       \\
 Gain  calibrator                                                                   &    J0239$-$0234                   \\             
 Flux density calibrator                                                        &     3C\,48                   \\     
 Bandpass calibrator                                                             &     3C\,48                   \\    
Array configuration                                                                &     B                   \\         
Projected baselines                                                                 &     0.2--11\,km                   \\
Primary beam                                                                        &    7.3 arcmin (FWHM)  at 6\,GHz                  \\
Synthesized beam size                                                          &       $1.5''\times 1.1''$ (PA, $16^{\rm \circ}.2$)                 \\
Map noise level                                                                      &    1.1\,$\mu$Jy\,beam$^{-1}$     \\
 \tableline 
\end{tabular} 
 \end{center}
\end{table}

\subsection{Method}
\label{photoz}

We estimate the radio/submm photometric redshifts of
ASXDF1100.053.1 and ASXDF1100.231.1 by fitting dust SED templates
to ALMA 1100-$\mu$m, SCUBA-2 850-$\mu$m and (J)VLA 6- or
1.4-GHz flux densities.

When making radio/(sub)mm photometric redshift estimates,
obtaining strong constraints around the peak of the dust SEDs is
important, to exclude spurious SED models which return dubious
redshift estimates due to the degeneracy between redshift and
dust temperature \citep[e.g.][]{bla02}. For most SMGs at
$z\approx 2$--3, the {\it Herschel} SPIRE images at 250, 350 and
500\,$\mu$m cover the redshifted dust SED peak. ASXDF1100.053.1
and 231.1 are not detected in the {\it Herschel} SPIRE maps
(Figs~\ref{fig:stamp53} and \ref{fig:stamp231}) and we have
therefore included $3\sigma$-upper limits from the SPIRE data at
250, 350 and 500\,$\mu$m as survival functions \citep{iso86}, as
was done for SCUBA 450-$\mu$m upper limits in radio/submm
photometric redshift estimates in \citet{are07}. 
Survival function enables us to derive redshift probability densities in the entire of a redshift range avoiding drastic changes due to upper limits in fluxes. 

Radio/submm photometric redshifts typically have larger
uncertainties than optical/near-IR photometric redshifts because
of the lack of clear SED features, such as continuum breaks.
Since redshift estimates using radio/(sub)mm colors
depend on the adopted dust SED, we need to use SEDs
representative of our target population, i.e.\ galaxies with
similar IR luminosities and similar redshifts.

Here, we adopt the SED template made from of 99 ALMA-identified SMGs,
derived from deep {\it Herschel} and ALMA submm and VLA radio
data presented in \citet{swi14}. The SEDs of ALMA-identified SMGs were
fitted with a library of 185 SEDs from \citet{cha03}, \citet{rie09},
\citet{dra07}, \citet{ivi10} and \citet{car11}, adopting optical/near-IR photometric redshifts from
  \citet{sim14}. The dust temperature of the best-fit SED of each
ALMA-identified SMG is listed in the paper.  We picked SEDs randomly
from the parent SED library along with the dust temperature
distribution (19--52\,K) for the ALMA-identified SMGs
derived in \citet{swi14} \footnote{The reformatted
    SED templates with dust temperatures used in \citet{swi14} are
    distributed at
    http://astro.dur.ac.uk/\~{}ams/HSOdeblend/templates/}, and
calculated the redshift probability density distribution for each
chosen SED. We bootstrapped this process and combined the derived
probability density distributions in order to achieve a redshift
probability density distribution weighted by the likelihood of each
SED temperature.

The multi-variate Gaussian probability distribution, $\Phi$, for {\it k} colors, is given by 
\begin{multline}
\Phi(\bm{c}_i-\bm{c}_0) = \\
(2\pi)^{-k/2}|{\rm {\boldsymbol
    A}}^{-1}|^{1/2}\exp(-\frac{1}{2}(\bm{c}_i-\bm{c}_0)'{\rm
{\boldsymbol A}}^{-1}(\bm{c}_i-\bm{c}_0)) \times \prod_{\nu} Surv,  \label{eq:1}
\end{multline}
where ${\rm {\boldsymbol A}}$ is a covariance matrix.  Here we assume
that any non-diagonal elements in the covariance matrix are zero,
therefore
$(\bm{c}_i-\bm{c}_0)'{\rm {\boldsymbol A}}^{-1}(\bm{c}_i-\bm{c}_0)$
can be substituted by standard $\chi^2$.  $Surv$ is a survival
function \citep{iso86}.  The survival function is expressed using an
complementary error function as
\begin{multline}
Surv = \frac{1}{\sqrt{2\pi}} \int^{\infty}_{(c_i(\lambda)-c_{obs}(\lambda))/\sigma_{obs}} e^{-t^2/2} dt. 
\end{multline}

We assume that the flux density errors follow Gaussian
distributions. The final redshift probability distribution, $P(z)$, of
any galaxy is the sum of the individual probabilities from the SEDs,
or explicitly
\begin{equation}
P(z) = a \sum_{i,\forall z} \Phi(\bm{c}_i-\bm{c}_0), 
\end{equation}
where $a$ is the normalization constant, such that
$\int^{z_{max}}_0 P(z)=1$ where $z_{max}=10$.  The asymmetric error
bars ($z_{-}$,$z_{+}$) correspond to 68\% confidence levels such that
$\int^{z_{+}}_{z_{-}} P(z)$\,d$z=0.68$ and ($z_{+}-z_{-}$) is
minimized. These calculations follow the methodology presented in
\citet{hug02} and that on the survival function.

\subsection{Resulting redshift estimates}
  
\begin{figure*}
\epsscale{1.0}
\plotone{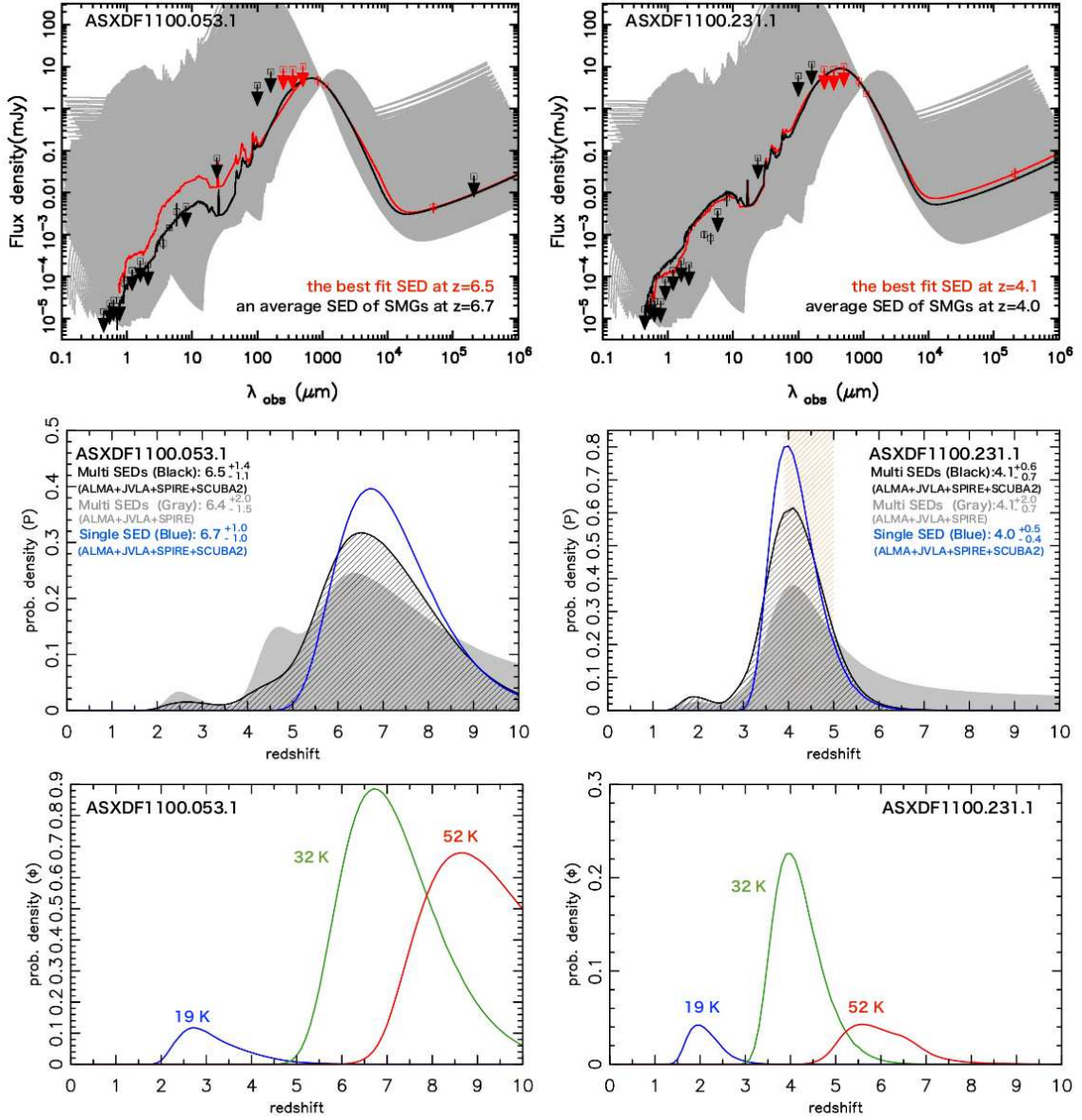}
\caption{Radio/(sub)mm photometric redshift of ASXDF1100.053.1
  and 231.1.  {\it Top:} Observed photometric data and model SEDs.
  Red open squares mark photometric data used in our
  radio/(sub)mm photometric redshift estimates: JVLA 6-GHz,
  ALMA 1100-$\mu$m, SCUBA2 850-$\mu$m, and upper limits in the SPIRE
  bands.  Black open squares mark photometric data not used in our
  redshift estimation.  The red line is the best-fit SED at the
  best-fit redshift in photometric redshift estimation.
  The grey-shaded area marks a range of all fit-SEDs at
    all redshifts. The black line represents the averaged SED of
    ALMA-identified SMGs at the best-fit redshift presented in
    \S\,\ref{sec:cross}. {\it Middle:} Redshift probability density
    distributions of radio/(sub)mm photometric redshift. The
    black hatched curve marks the redshift probability density
    distribution. The grey curve shows that without the SCUBA2
    850-$\mu$m data. The blue line marks that of a single SED
    template, the average SED of ALMA-identified SMGs. The derived
    photometric redshift for each estimate is displayed in the
    panels. The orange hatched area marks a redshift range where the
    mid-IR color of ASXDF1100.231.1 is explained by the redshifted
    H$\alpha$ emission line in the IRAC 3.6-$\mu$m band, as
    discussed in \S\ref{sec:dis2}. {\it Bottom:}
    Probability densities ($\Phi$) for individual SEDs of 19, 32 and
    52\,K using all (sub)mm/radio bands which give the
    contributions of different $T_d$ temperature to the combined
    photometric redshift. }
\label{fig:phz_053.1}
\end{figure*}

The radio/submm photometric redshift probability distributions
for ASXDF1100.053.1 and 231.1 are shown with black curves in
Fig.~\ref{fig:phz_053.1} along with fit-SEDs. 
The resultant photometric redshifts for ASXDF1100.053.1 and 231.1 are $6.5^{+1.4}_{-1.1}$ and  $4.1^{+0.6}_{-0.7}$, respectively.

Probability densities, $\Phi$, for each SED with $T_d=$19,
32 and 52\,K given by Equation~\ref{eq:1}, are shown at the bottom of
Fig.~\ref{fig:phz_053.1}, with the aim of understanding the
contributions of these SEDs to the combined photometric redshift
probability distributions. The $\Phi$ density plots indicate that the
low probabilities of low-redshift solutions in the combined redshift
probability distribution, $P(z)$, are due to two factors: (1) the
rarity of cold SEDs due to the dust temperature distribution, and (2)
the cold SEDs give poor fits. The $\Phi$ plots also demonstrate that
solutions for cold SEDs are less plausible for the two SMGs,
regardless of the rarity or otherwise of cold SEDs. 

\begin{figure*}
\epsscale{1.1}
\plotone{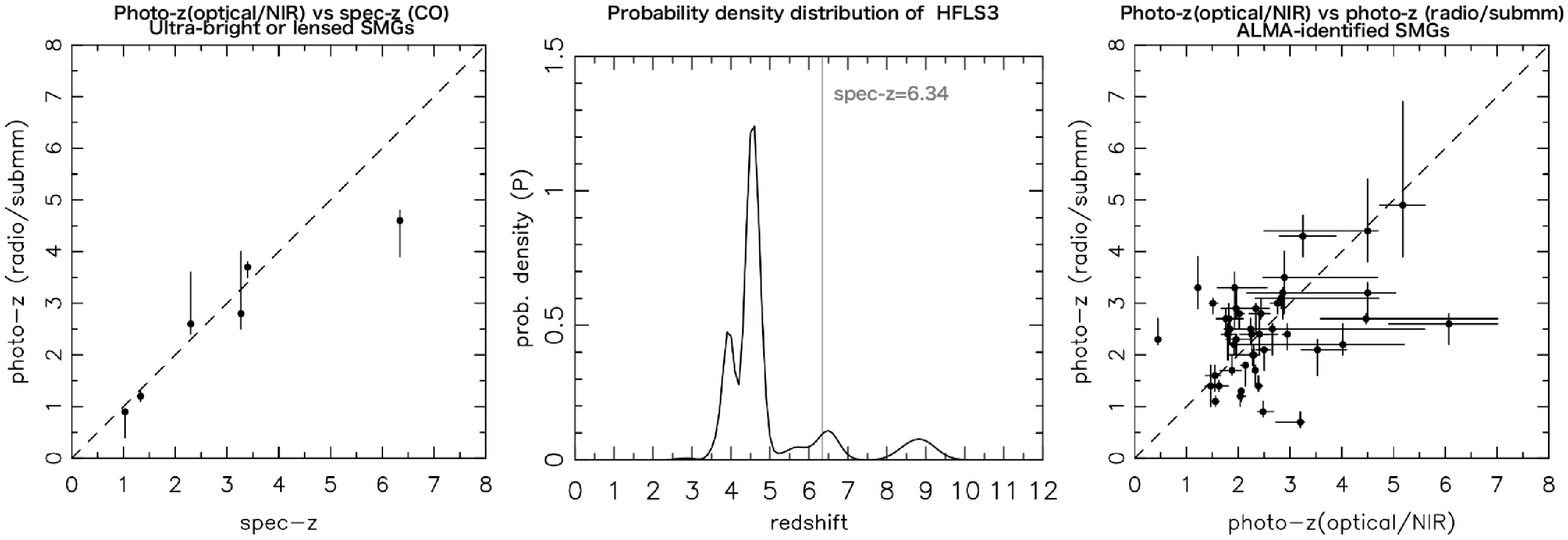}
\caption{Radio/(sub)mm photometric redshift estimation.  {\it
    Left:} Comparison of radio/(sub)mm photometric redshift
  with spectroscopic redshift obtained via CO for six bright or lensed
  SMGs from the literature \citep{ivi10, ika11, rie13, war13, mes14}.
  {\it Middle:} Redshift probability density distribution for HFLS3,
  shown in order to explain what happens when estimating its redshift
  using radio/(sub)mm photometry.  {\it Right:} Comparison of
  radio/(sub)mm photometric redshifts with optical/near-IR
  photometric redshifts for 46 ALMA-identified SMGs with radio
  detections \citep{sim14, swi14}.  }
\label{fig:bench}
\end{figure*}

\subsection{Benchmark tests of the redshift estimates}
 
It is informative to perform some benchmark tests, using SMGs
with known spectroscopic redshifts to assess whether our method
returns sensible values and to evaluate systematics in our photometric
redshift estimates. We have found seven bright or lensed SMGs with CO
spectroscopic redshifts which have SPIRE, $\sim 1000$-$\mu$m and radio
photometry in the literature \citep{ivi10, ika11, rie13, war13,
  mes14}. Fig.~\ref{fig:bench} shows comparisons of their
radio/(sub)mm photometric redshifts and their spectroscopic
redshifts. All except HFLS3 show good agreement. The under-estimation
of the redshift when using the radio/(sub)mm method for HFLS3
can be explained by its abnormally high dust temperature (56\,K). As
the probability density distribution shows (see middle in
Fig.~\ref{fig:bench}), there is a small local peak around the
spectroscopic redshift with a $T_d$ similar to that of HFLS3.

In addition to benchmark tests with a spectroscopic sample, we also
conducted another benchmark test using 46 radio-detected
ALMA-identified SMGs from ALESS with optical/near-IR photometric
redshifts \citep{sim14, swi14}. ALESS sources were originally
880-$\mu$m-selected SMGs and are expected to be drawn from the same
population as ASXDF1100.053.1 and 231.1. We estimate radio/submm
photometric redshifts using SPIRE 250--500-$\mu$m, ALMA 880-$\mu$m and
VLA 1.4-GHz flux densities \citep[see these flux densities in Table~A1
of][]{swi14}. A comparison of their radio/submm-estimated photometric
redshifts and optical/near-IR estimates is shown in
Fig.~\ref{fig:bench}.  We derived
$\Delta z=(z^{\rm radio}_{\rm photo}-z^{\rm opt}_{\rm
  photo})/(1+z^{\rm opt}_{\rm photo})$. Its median and 1$\sigma$
dispersion are $-$0.01 and 0.27, respectively. We should
note that there is no contamination between optical/near-IR
photometric redshifts of $\lesssim4$ and radio/(sub)mm
photometric redshifts of $\gtrsim5$.
 
These benchmark tests suggest that our radio/(sub)mm
photometric redshifts using multi-SEDs do not suffer strong
systematics. 

\subsection{Cross-checking photometric redshifts}
\label{sec:cross}

We first derived photometric redshifts using the average SED of 99
ALMA-identified SMGs from \citet{swi14}. This redshift estimate is
expected to give us the most reliable redshift for typical SMGs but
will under-estimate the uncertainty due to the plausible diversity of
SEDs. The derived redshift probability density distributions for
ASXDF1100.053.1 and 231.1, based on the average SED, show results
consistent with radio/submm-based photometric redshifts
(Fig.~\ref{fig:phz_053.1}): The respective photometric redshifts based
on the average SED are $6.7^{+1.0}_{-1.0}$ and $4.0^{+0.5}_{-0.4}$.

Redshift probability densities based on our redshift estimates without
SCUBA-2 850-$\mu$m data are shown in Fig.~\ref{fig:phz_053.1}. The
850-$\mu$m detection allows a smaller uncertainty and sharpens the
redshift probability densities. This implies that the model 850-$\mu$m
flux densities of the SMGs based on photometric redshifts from ALMA
1100-$\mu$m and (J)VLA radio colors and the upper limits at SPIRE
bands are consistent with the observed 850-$\mu$m flux densities. 

\subsection{Possible effects of the CMB on redshift estimation}

In the very early Universe, the cosmic microwave background (CMB)
can have effects on observed submm and radio flux
densities \citep[see e.g.][]{zha16}. Here we discuss possible
contributions of the CMB to the radio/submm photometric
redshifts using toy models for CMB effects.

On the basis of the predictions of CMB effects on observed
total submm flux densities (\citealt{dac13} deals with the
effect on total flux densities, and \citealt{zha16} explores the
spatially resolved effects), we took into account two CMB effects on
observed submm flux densities: the effect on intrinsic
far-IR/submm dust SEDs, and on the detectability of SMGs
against the CMB background.  We evaluate these effects on the observed
flux densities at 1100, 850, 500, 350 and 250\,$\mu$m for the
$T_{\rm d}$ of each SED, in the same manner as
\citet{dac13}. 

Observed radio flux densities of distant galaxies are expected to
get fainter as a function of redshift due to suppression of
synchrotron emission by inverse Compton (IC) losses off the CMB
\citep{mur09}. The suppression off radio flux densities by the
CMB depends on the strength of the magnetic field ($B$) in a
galaxy, about which we know very little.

On the basis of observational studies of SMGs in the literature,
\citet{mur09} suggested that SMGs can have a strong $B$,
potentially $\gtrsim300$\,$\mu$G. \citet{mcb14} reported a
minimum $B$ strength of $\gtrsim150$--500\,$\mu$G for local
(U)LIRGs, based on observed synchrotron flux densities. They
expected a stronger $B$, $>600$\,$\mu$G, based on measurements of
Zeeman splitting in OH masers. Given that ASXDF1100.053.1 and
231.1 have compact mm-wave sizes and surface IR
luminosity densities similar to those of local ULIRGs (see
\S\,\ref{sec:size} and \S\,\ref{sec:mmp}), these studies also
support a strong $B$ for our sample.

In this paper, we investigate how the effect of the CMB on radio
emission contributes to radio/mm photometric redshift
estimates where $B=100$ and 300\,$\mu$G: 300\,$\mu$G is taken as the
value for SMGs, and 100\,$\mu$G is used to examine what happens if the
magnetic field is weaker. We determined the predicted suppression of
non-thermal emission by the CMB using the equations and assumptions
provided in \citet{mur09}: we modeled the synchrotron emission 
by subtracting free-free and dust emission, where we model free-free
emission from $L_{\rm IR}$ based on the equation (16) in the
literature. 

\begin{figure*} \epsscale{1.2} \plotone{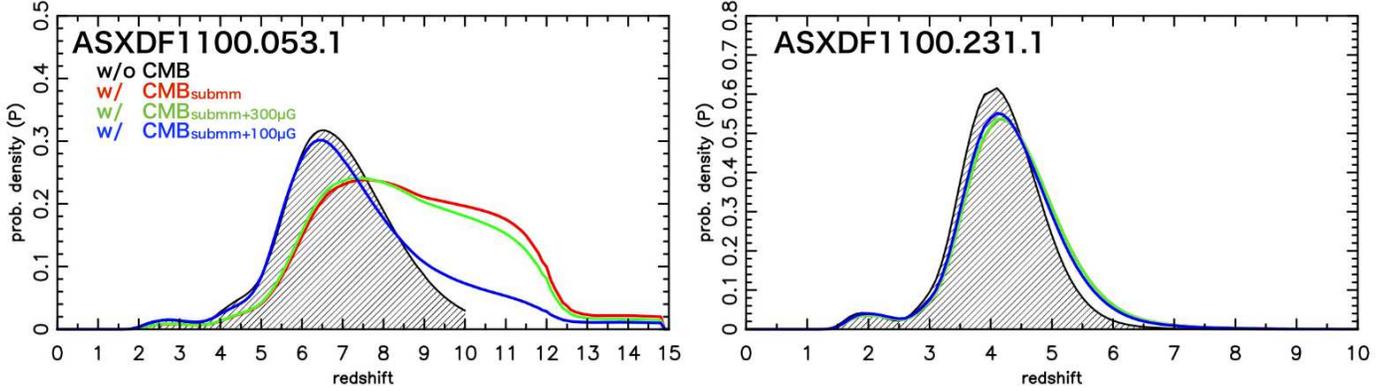}
  \caption{Redshift probability densities of
    radio/(sub)mm photometric redshift estimates when
    including the effects of the CMB for ASXDF1100.053.1 and
    231.1. The grey curve is a probability density distribution
    without any CMB effects, i.e.\ it is the same plot shown in
    Fig.~\ref{fig:phz_053.1}. The red curve shows a probability
    density distribution with the CMB effect on observed
    (sub)mm flux densities. The green curve is a probability
    density distribution with the CMB effects on both 
    (sub)mm and radio flux densities for $B=300$\,$\mu$G. The blue
    curve is for $B=100$\,$\mu$G. The redshift
      range is extended to $z=15$ for ASXDF1100.053.1 due to the
      large probability of $z\geq10$. For the direct comparison
      with the result shown in Fig.~\ref{fig:phz_053.1}, the
      probability densities are scaled in $z=0$--10.} 
  \label{fig:probcmb}
\end{figure*}

Fig.~\ref{fig:probcmb} shows the resulting redshift probability
density distribution, including the CMB effects, for both
ASXDF1100.053.1 and 231.1: (1) probability density without any CMB
effects; (2) with the CMB effect at (sub)mm wavelengths; (3)
with both of the CMB effects ($B=300$\,$\mu$G); (4) with both of the
CMB effects ($B=100$\,$\mu$G). We see that taking the effects of the 
CMB into account pushes the photometric redshifts to higher values.
However, as we do not have spectroscopic redshifts for the two SMGs, we cannot determine how strong this effect really is.

\section{ALMA mm-wave source sizes}\label{sec:size}

The first millimetric size measurements of ASXDF1100.053.1 and 231.1
were determined using our ALMA Cycle-1 data, which covered up to a
$uv$ distance of $\sim 400$\,k$\lambda$, with a synthesized beam size
of $\sim 0''.70$ \citep[FWHM -- ][]{ika14}. These visibility data
assumed Gaussian profiles and suggested compact millimetric sizes:
$0.28^{+0.04}_{-0.04}$ and $0.12^{+0.08}_{-0.08}$ arcsec (FWHM) for
ASXDF1100.053.1 and 231.1, respectively. In this section, we re-assess
their millimetric sizes, combining Cycle-1 and -2 data, which now
cover up to 1500\,k$\lambda$.

\begin{figure*}
\epsscale{1.2}
\plotone{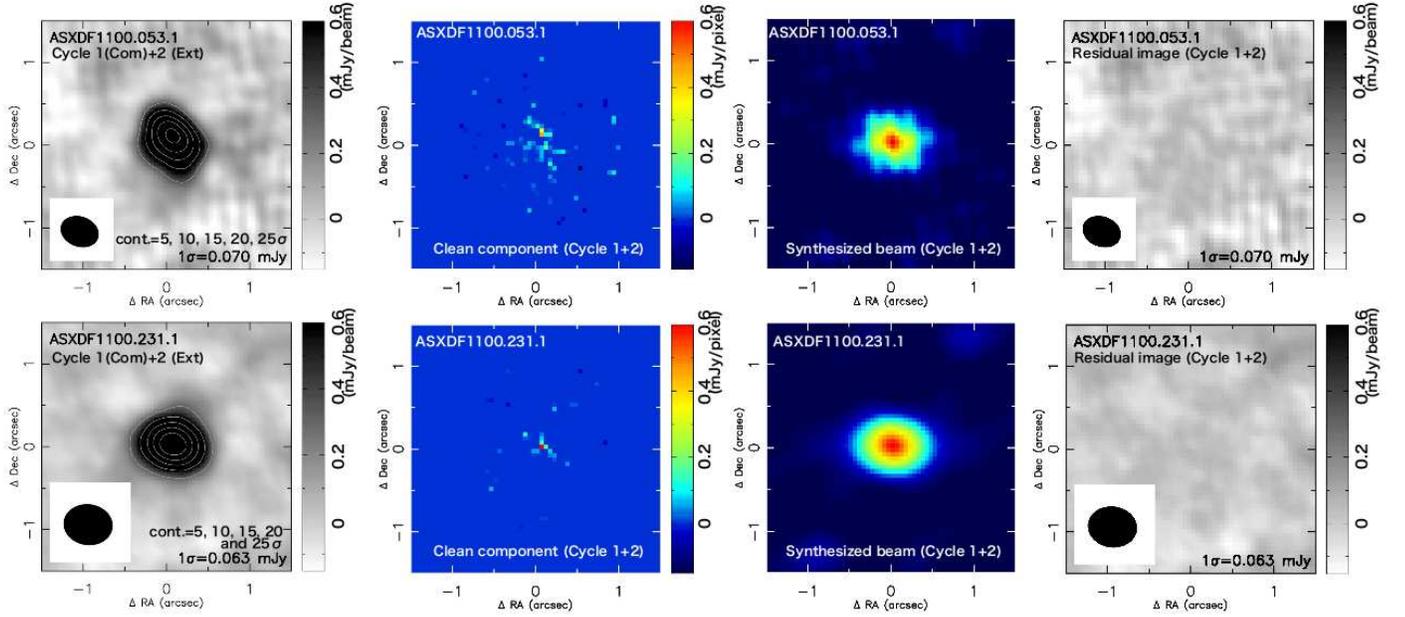}
\vspace{2mm}
\caption{ALMA images of ASXDF1100.053.1 and 231.1.  {\it 1st column:}
  Cleaned ALMA 1100-$\mu$m continuum images taken in ALMA Cycles 1 and
  2.  Synthesized beams of the combined data are
  0$''$.46$\times$0$''$.35 (PA, 69$^{\circ}$) and
  0$''$.57$\times$0$''$.48 (PA, 82$^{\circ}$) for ASXDF1100.053.1 and
  231.1, respectively.  Contours are shown at 5, 10, 15, 20 and
  25$\sigma$.  The flux density unit is mJy\,beam$^{-1}$.  The r.m.s.\
  noise level is shown at bottom in each panel.  {\it 2nd column:}
  Clean component maps of the combined ALMA images are shown in the
  middle panel. The clean component maps were obtained by cleaning
  down to 1$\sigma$.  {\it 3rd column:} Synthesized beams for the
  combined ALMA images.  {\it 4th column:} Residual maps after
  subtracting the clean components convolved with the synthesized
  beams.  The pixel scale is 0$''$.05 in all images.
}  \label{fig:almaimages}
\end{figure*}

In Fig.~\ref{fig:almaimages}, ALMA maps for ASXDF11100.053.1 and
231.1 are shown. These maps were generated from the combined ALMA
1100-$\mu$m data, cleaning down to the 1-$\sigma$ depth in a
circle with a radius of 1 arcsec using the {\sc clean} task in
CASA. The pixel scale is 0$''$.05 pixel$^{-1}$.

\subsection{Mm-Wave size measurements in visibility data}
 
\begin{figure*} \epsscale{1.2} \plotone{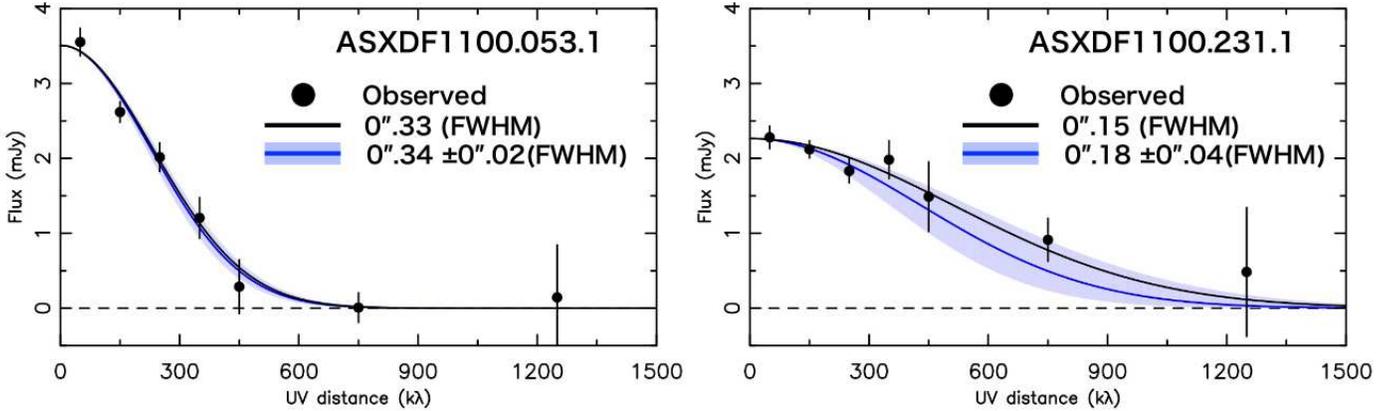} \caption{ALMA
    $uv$-distance versus amplitude plots of ASXDF1100.053.1 and
    231.1. Black solid points are the observed data. Binning sizes in
    $uv$-distance are 100\,k$\lambda$ out to 500\,k$\lambda$ and
    500\,k$\lambda$ between 500--1500\,k$\lambda$. The black line is a
    $uv$-amp model of the best-fit Gaussian component. The blue line
    and shaded area are possible solutions for the corrected source
    size, with errors based on Monte Carlo simulations.}
  \label{fig:uvamp} \end{figure*}

First, we measure mm-wave sizes of ASXDF1100.053.1 and 231.1 with
the ALMA visibility data, in the same manner as \citet{ika14}. We use
$uv$-distance versus amplitude plots (hereafter $uv$-amp plots) for
the measurements (Fig.~\ref{fig:uvamp}). Modelling sources with
$uv$-amp plots helps us to avoid underestimating their flux densities,
since we can interpolate/extrapolate across incomplete visibility
coverage. We assume symmetric Gaussian profiles, as is usually done in
the literature. Circularized effective radii estimated using $uv$
plots are useful, even for sources with asymmetric profiles
\citep{ika14}. Bin sizes adopted in $uv$-distance are 100\,k$\lambda$
out to 500\,k$\lambda$ and 500\,k$\lambda$ between
500--1500\,k$\lambda$. The estimated sizes of ASXDF1100.053.1 and
231.1 are then 0$''$.33 and 0$''$.15 (FWHM), respectively
(Fig.~\ref{fig:uvamp}). Correcting these `raw' mm-wave sizes for
systematic effects using Monte Carlo simulations, the mm-wave sizes
are then 0$''$.34$^{+0.02}_{-0.02}$ and 0$''$.18$^{+0.04}_{-0.04}$ for
ASXDF1100.053.1 and 231.1, respectively, consistent with our previous
measurements.

\subsection{Mm-Wave size measurements in clean component maps} 

Next, we derive $R_{\rm c,e}$ for ASXDF1100.053.1 and 231.1 using ALMA
clean component maps, as shown in Fig.~\ref{fig:almaimages}.  These
maps were generated from the combined ALMA 1100-$\mu$m data by running
the {\sc clean} task in CASA. 
Our motivation is to measure $R_{\rm c,e}$ directly,
without any assumed model, exploiting the high signal-to-noise ratios
of $\sim30$. 

\begin{figure*}
\epsscale{1.2}
\plotone{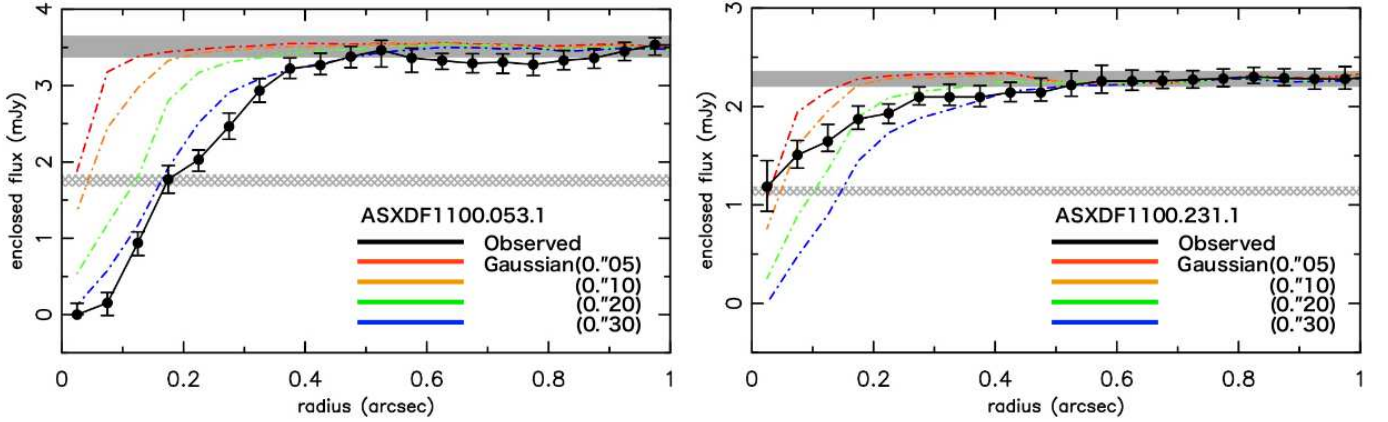}
\caption{Enclosed flux densities as a function of radius measured in
  the clean component maps for ASXDF1100.053.1 and 231.1.  The black
  solid line shows observed data.  Colored dot-dashed lines show
  enclosed flux densities of models with Gaussian profiles with
  various sizes from 0$''$.05 to 0$''$.30 (FWHM).  The enclosed flux
  density profiles of Gaussian models are reproduced in the same
  manner using the actual data, and taken from simulations presented
  in \S\ref{sec:appendix}.  Filled grey lines show total ALMA flux
  densities with errors measured independently in the beam-convolved
  ALMA images, which are listed in Table~\ref{tbl-1}.  Hatched grey
  lines show half of the total ALMA flux densities with the errors for
  finding half-light radii.}  \label{fig:encflux}
\end{figure*}

Fig.~\ref{fig:encflux} shows enclosed flux densities as a function of
radius for ASXDF1100.053.1 and 231.1, measured in the clean component
maps.  Total flux densities of the two SMGs in the clean component
maps are consistent with the fluxes measured in the beam-convolved
ALMA continuum images listed in Table\,2, despite the potential absence of any
$<1\sigma$ components.  Flux density errors are estimated based on
Monte Carlo simulations using 100 independent sets of visibility data
generated from the actual ALMA data.  In the simulations, we input a
Gaussian model with the same flux as one of the real sources, then
imaged these data, generating clean component maps.  We then measure
the enclosed flux densities in the same manner as that for the real
sources.  We repeated this process with source sizes between 0$''$.025
and 0$''$.800 in steps of 0$''$.025 (FWHM) to reconstruct observed
enclosed flux densities in each bin.  We adopted a flux density error
in the simulation with an enclosed flux density closest to a real
measured flux density in each radius bin as the error for the real
measurements.  We refer the reader to \S\ref{sec:appendix} where we
describe the simulations in more detail.
 
For ASXDF1100.053.1, based on the obtained enclosed flux density plot
and the total flux density, we determine $R_{\rm c,e}$ of
0.17$^{+0.02}_{-0.01}$ arcsec.  Since the half width half maximum of a
symmetric Gaussian is equivalent to $R_{\rm c,e}$, the size obtained
from the clean component map is consistent with that from the $uv$-amp
plot.

For ASXDF1100.231.1, the flux density in the center pixel is
$1.19^{+0.27}_{-0.24}$\,mJy\,beam$^{-1}$.  This corresponds to
$52^{+7}_{-6}$\% of its total flux density.  From the obtained
enclosed flux density plot, with linear interpolation, we find
$R_{\rm c,e}=0.025^{+0.015}_{-0.00}$ arcsec, meaning the half light
radius of ASXDF1100.231.1 is $\leq 0''$.04.  $R_{\rm c,e}$ determined
via the clean component map is approximately $2\times$ smaller than
that determined from the $uv$-amp plot, $R_{\rm c,e}=0.09$ arcsec.

The enclosed flux density plot suggests that ASXDF1100.231.1 cannot be
modeled with single Gaussian profile: ASXDF1100.231.1 appears to
comprise a compact, intense mm emitting region, located in its
center region, and a fainter, extended region.  The different
$R_{\rm c,e}$ values determined using the $uv$-amp plot and the clean
component map can thereby be understood.
 
Monte Carlo simulations of source size measurements in ALMA clean
component maps are described in Appendix~\ref{sec:appendix}.  According to
these simulations, this method of measuring source sizes is useful
down to $R_{\rm c,e}=0''$.025.  The simulations show that the measured
sizes of ASXDF1100.053.1 and 231.1 are not expected to suffer large
systematic errors.

\section{On the nature of ASXDF1100.053.1 and 231.1}
\label{sec:dis2}

Here we determine the properties of ASXDF1100.053.1 and 231.1 from
multi-wavelength data, adopting the radio/submm photometric
redshifts.  We discuss the possible role of the two SMGs, which are
faint in the {\it Herschel} bands, and at optical/near-/mid-IR and
radio wavelengths, in the context of galaxy evolution.

\subsection{Optical/near-IR SED fitting}
\label{sec:optnir}

In order to characterize the optical/near-IR properties of
ASXDF1100.053.1 and 231.1, we conducted an SED-fitting analysis across
optical--mid-IR wavelengths.  We adopted a Chabrier IMF \citep{chab03}
and stellar population synthesis models by \citet{bru03}.  Dust
extinction was considered according to the prescription by
\citet{cal00}.  We adopted metallicity, $Z_{\rm \odot}=0.02$.  We
performed the analysis using the code, {\it Le Phare}
\citep{arn99,ilb06}.  These SMGs are detected in only three or four
filters in the available optical/mid-IR broad band images, similar to
the extremely red mm source analyzed by \citet{cap14}. We
derived optical/near-IR properties at fixed redshifts of $z=5.5$, 6.5
and 7.5 for ASXDF1100.053.1, and $z=3.5$, 4.5 and 5.5 for
ASXDF1100.231.1, to reduce parameter space.  These values span a range
of approximately $\pm1$ around the best radio/submm
photometric redshifts.  We conducted the SED fitting both with and
without considering emission lines.  Free parameters were
star-formation history, age and dust extinction.

\begin{figure}
\epsscale{1.2}
\plotone{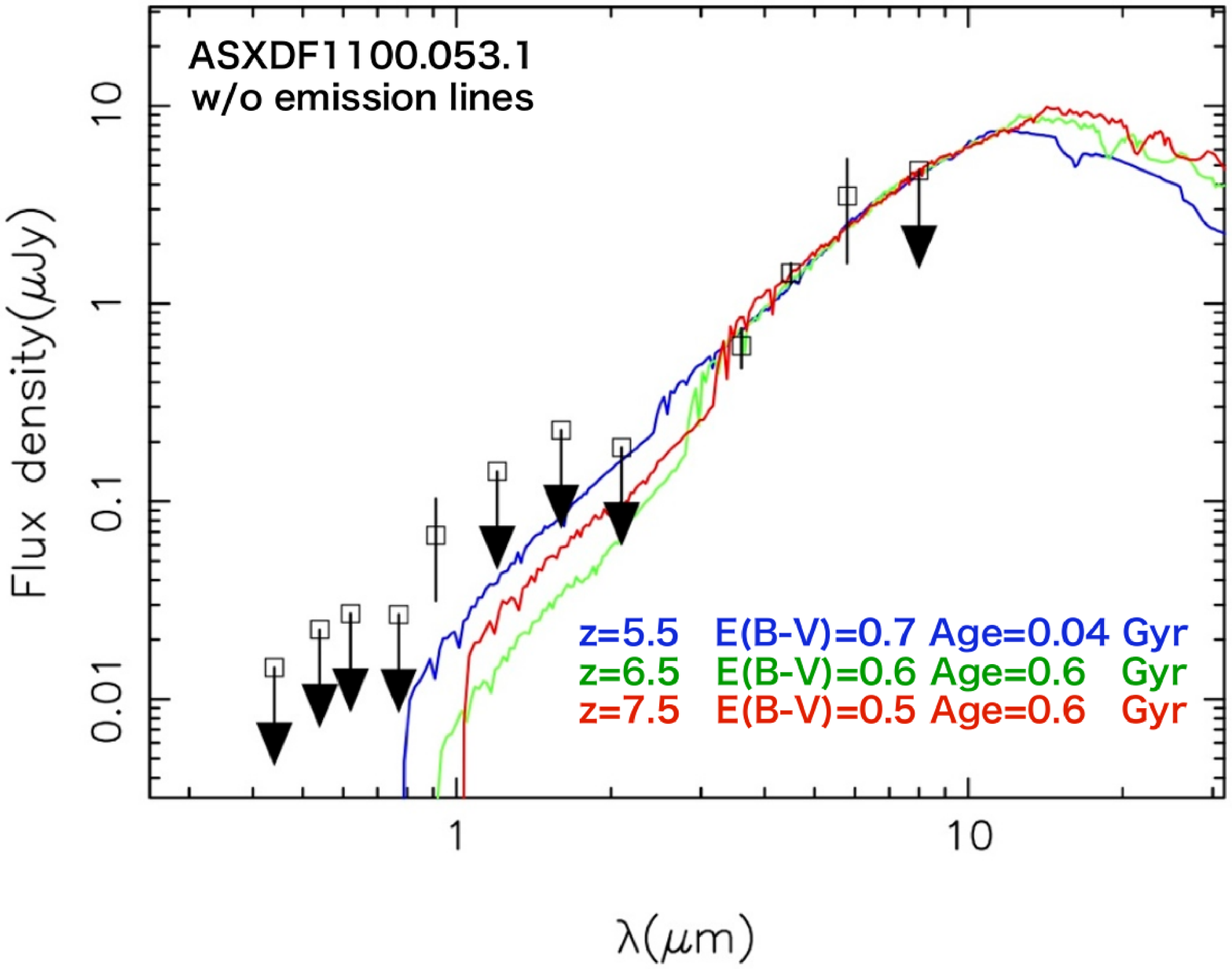}
\plotone{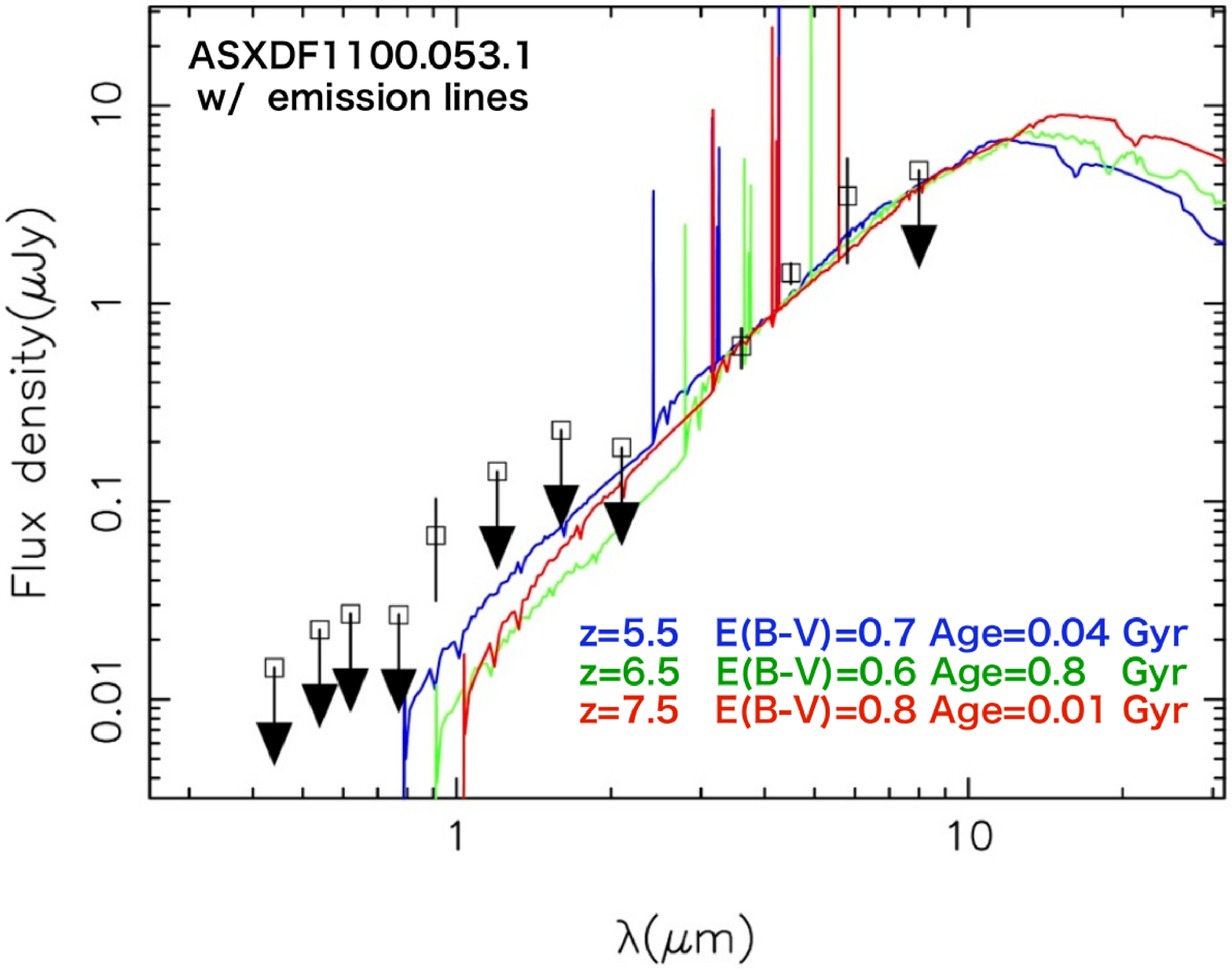}
\caption{ASXDF1100.053.1 SED at observed optical/near-/mid-IR
  wavelengths.  {\it Top:} Best-fit SEDs, ignoring emission lines,
  for $z=$5.5, 6.5 and 7.5, as obtained with {\it Le Phare}.  Black
  points correspond to the observed flux densities, while arrows
  indicate 2$\sigma$ upper limits.  {\it Bottom:} Best-fit SEDs, this
  time including emission lines.  }
\label{fig:nirsed053}
\end{figure}

\begin{figure}
\epsscale{1.2}
\plotone{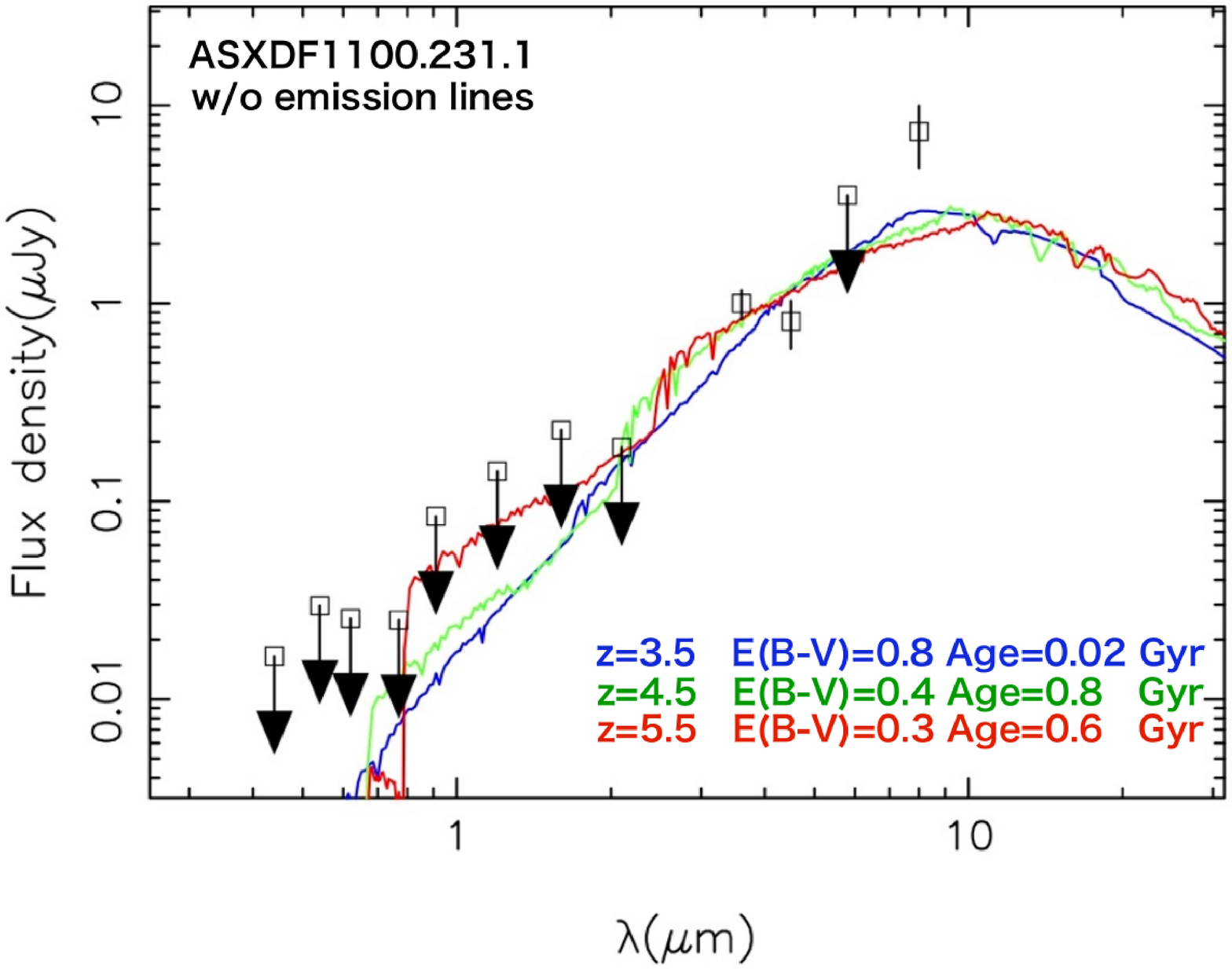}
\plotone{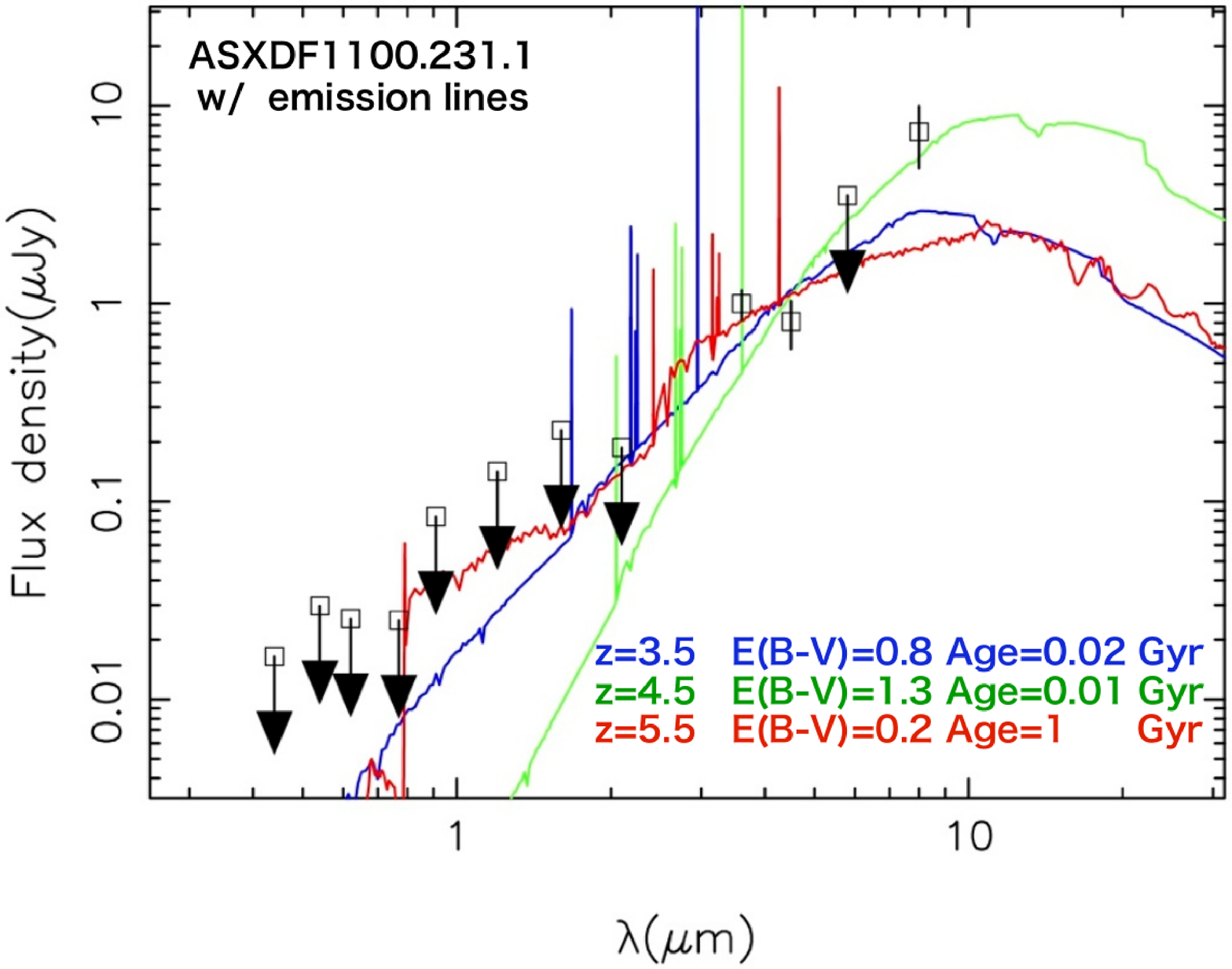}
\caption{ASXDF1100.231.1 SED at observed optical/near-/mid-IR
  wavelengths.  {\it Top:} Best-fit SEDs, without considering emission
  lines, toward ASXDF1100.231.1 for $z=$3.5, 4.5 and 5.5, as obtained
  with {\it Le Phare}. Black points correspond to the observed flux
  densities, while arrows indicate 2$\sigma$ upper limits.  {\it
    Bottom:} Best-fit SEDs, this time including emission
  lines.}
\label{fig:nirsed231}
\end{figure}

The best-fit SEDs are shown in Figs~\ref{fig:nirsed053} and
\ref{fig:nirsed231} and the derived parameters are summarized in
Table~\ref{tbl:prop}.  The observed optical/mid-IR SED of
ASXDF1100.053.1 is well fit at $z=5.5$, 6.5, and 7.5, both with and
without emission lines as the minimum $\chi^{2}$s show.
ASXDF1100.053.1 has an extremely red color, $[3.6]-[4.5]=0.9$, which
can be reproduced by heavy dust extinction, $E(B-V)\sim 0.6$--0.8.  In
all cases, the observed $z'$-band flux is not reproduced, but the
detection at $z'$-band is only marginal ($2\sigma$).  If the detection
is real, the rest-frame UV light may come from less obscured regions
in ASXDF1100.053.1.  It is worth mentioning that the Ly$\alpha$ line
should fall in the $z'$-band at $z=6.0$--7.2, but the importance of
this line in emission at such redshifts is still unclear.

The observed optical/mid-IR SED of ASXDF1100.231.1 is not well fit by
the model SEDs.  Its unusual colors, $[2.2]-[4.5]<1.7$,
$[3.6]-[4.5]=-0.1$ and $[4.5]-[8.0]=2.3$, imply a possible excess in
the IRAC 3.6-$\mu$m band.  Remarkably, the SED of ASXDF1100.231.1 is
best fit by the model at $z=4.5$, including emission lines, with the
heaviest extinction among our SED fits. The H$\alpha$ emission line
enters the IRAC 3.6-$\mu$m band at $z=3.9$--5.0, consistent with the
radio/submm photometric redshift, $4.1^{+0.6}_{-0.7}$
(Fig.~\ref{fig:phz_053.1}).  Interestingly, two of 77 ALMA-identified
SMGs in \citet{sim14}, ALESS1.2 and 65.1 have a similar color and
excesses in the 3.6-$\mu$m band ($[3.6]-[4.5]<0$, $[2.2]-[4.5]>0$ and
$[4.5]-[8.0]>0$) and both have $z_{\rm photo-z, opt}=4.65^{+2.34}_{-1.02}$
\citep{sim14} and $z_{\rm spec}=4.4445\pm0.0005$
\citep{swi12}, respectively.

The stellar mass of ASXDF1100.053.1, derived from the SED fitting
without emission lines, would be $[1.0-1.6]\times10^{11}$\,M$_{\odot}$
at $z=5.5$--7.5.  When taking emission lines into account, the stellar
mass becomes $[0.8-1.7]\times10^{11}$\,M$_{\odot}$.  For
ASXDF1100.231.1: ignoring emission lines yields
$[0.5-3.3]\times10^{10}$\,M$_{\odot}$ at $z=3.5$--5.5; with emission
lines, $[0.4-2.8]\times10^{10}$\,M$_{\odot}$.  These stellar masses,
which are summarized in Table~\ref{tbl:prop}, are consistent with
those of known ALMA-identified SMGs \citep[e.g.][]{dac15}.

\subsection{Mm properties}\label{sec:mmp}

The IR luminosities ($L_{\rm IR}$, rest-frame 8--1000\,$\mu$m)
and star-formation rates (SFRs) of ASXDF1100.053.1 and 231.1, can be
estimated from the ALMA 1100-$\mu$m continuum, having adopted the
average SED of ALMA-identified SMGs from \citet{swi14}, using
\begin{equation}
{\rm SFR} ({\rm M_{\odot}\,yr^{-1}}) = 1.0\times10^{-10} L_{\rm IR} ({\rm L_{\odot}})
\end{equation} 
for a Chabrier IMF \citep{chab03}, according to the formula provided by \citet{ken98}. 

For ASXDF1100.053.1, $L_{\rm IR}=[4.8-6.0]\times10^{12}$\,L$_{\odot}$
at $z=5.5$--7.5 and the SFR is in the range
580--600\,M$_{\odot}$\,yr$^{-1}$.  These $L_{\rm IR}$ and SFR
estimates are largely independent of redshift, due to the strong
negative K-correction at submm wavelengths
\citep[e.g.][]{bla02}.  When we consider the CMB effects on the
observed 1100-$\mu$m fluxes, we find
$L_{\rm IR}=[6.5-7.4]\times10^{12}$\,L$_{\odot}$ and
an SFR between 650--740\,${\rm M_{\odot}}$\,yr$^{-1}$.

The corresponding values for ASXDF1100.231.1 are without the CMB
correction are $L_{\rm IR}=[3.8-4.3]\times10^{12}$\,L$_{\odot}$ at
$z=3.5$--5.5, with an SFR in the range
380--430\,${\rm M_{\odot}}$\,yr$^{-1}$; with CMB corrections:
$L_{\rm IR}=[4.2-4.5]\times10^{12}$\,${\rm L_{\odot}}$ and
420--450\,${\rm M_{\odot}}$\,yr$^{-1}$.  These values and errors are
listed in Table~\ref{tbl:prop}.

As it was revealed in \S\,\ref{sec:size}, ASXDF1100.053.1 and 231.1
have the compact mm-wave sizes.  From the clean component maps, we
find $R_{\rm c,e}=$0.17$^{+0.02}_{-0.01}$ arcsec for ASXDF1100.053.1,
and $\leq$0.04 arcsec for ASXDF1100.231.1.  Given physical scales at
$z=5.5$--7.5, the $R_{\rm c,e}$ of ASXDF1100.053.1 corresponds to
0.88--1.1\,kpc.  Given the physical scales at $z=3.5$--5.5, the
measured $R_{\rm c,e}$ for ASXDF1100.231.1 is $\leq$0.20--0.25\,kpc.
For further characterization of sizes, we also derived
$R_{\rm c, 0.9}$, the circularized radii that include 90\% of the
total flux density from the enclosed flux functions
(Fig.~\ref{fig:encflux}): $R_{\rm c,0.9}$ (median) of ASXDF1100.053.1
for $z=5.5$--7.5 is 2.2--2.6\,kpc and that of ASXDF1100.231.1 for
$z=3.5$--5.5 is 1.6--2.1\,kpc. We applied systematic corrections based on the simulations shown in \S\,\ref{sec:appendix}. 
The compact nature of ASXDF1100.231.1 is similar to the few $\times$100-pc clumps discovered by
0$''$.015--0.05$''$ imaging of SMGs by \citet{ion16} and
\citet{ote16}.  These $R_{\rm c,0.9}$ values are consistent with
suggestions of extended emission in SMGs in \citet{hod16} and \citet{ion16}. 

\begin{figure}
\epsscale{1.1}
\plotone{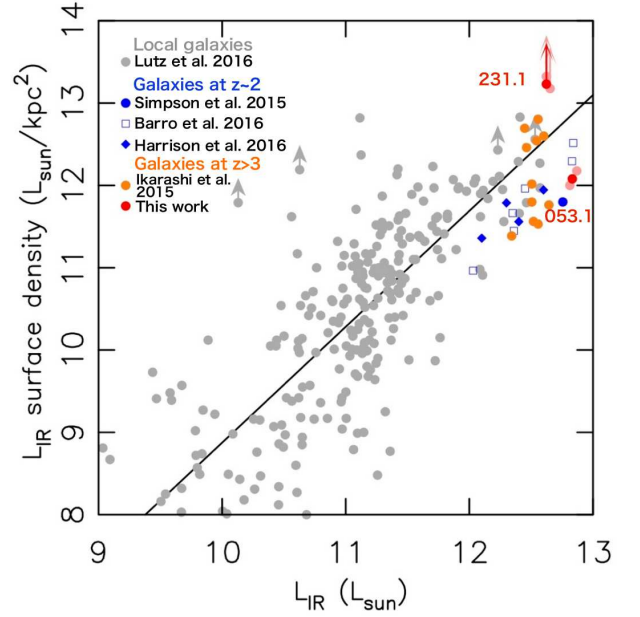}
\caption{IR luminosity versus surface IR luminosity
  density plots for known local and high-redshift galaxies.  Red dots
  mark ASXDF1100.053.1 at $z=6.5$ and 231.1 at $z=5.5$.  Pink dots
  mark ASXDF1100.053.1 at $z=5.5$ and 7.5, and 231.1 at $z=3.5$ and
  5.5, to show uncertainties due to redshift.  Grey dots mark local
  galaxies and a black solid line shows the fit-line for local
  galaxies presented by \citet{lut16}.  Blue markers show known
  $z\sim2$ galaxies: open squares mark so-called main-sequence
  galaxies \citep{bar16}, filled diamond X-ray AGNs \citep{har16}, and
  the filled circle marks the median for SMGs \citep{sim15a}.  Orange
  dots mark $z\gtrsim3$ candidate SMGs by \citet{ika14}.  The sizes of
  these high-redshift galaxies in the literature come from
  measurements using ALMA data.  $L_{\rm FIR}$ (40--120\,$\mu$m) was
  converted to $L_{\rm IR}$ (rest-frame 8--1000\,$\mu$m) based on an
  expected offset of 0.3 in log scale, based on the empirical
  far-IR/radio luminosity correlation \citep{yun01} and the
  IR/radio luminosity correlation \citep{bel03}.
} \label{fig:lumisize}
\end{figure}

Surface IR luminosity densities ($\Sigma_{L_{\rm IR}}$) based on
$R_{\rm c,e}$ and $L_{\rm IR}$ are
$[1.0-1.5]\times10^{12}$\,${\rm L_{\odot}}$\,kpc$^{-2}$ at
$z=5.5$--7.5 for ASXDF1100.053.1.  Those of ASXDF1100.231.1 are
$[1.5-2.1]\times10^{13}$\,${\rm L_{\odot}}$\,kpc$^{-2}$ for
$z=3.5$--5.5.  Fig.~\ref{fig:lumisize} shows $L_{\rm IR}$ versus
surface IR luminosity density for ASXDF1100.053.1 and 231.1, and
demonstrates that for $z\gtrsim4$ they are similar to local and
high-redshift galaxies from \citet{ika14,sim15a,bar16,har16,lut16}.
For $z\gtrsim6$ and $z\sim4$, respectively, ASXDF1100.053.1 and 231.1
have surface IR luminosity densities consistent with an
empirical $L_{\rm IR}-\Sigma_{L_{\rm IR}}$ relation
($\log (\Sigma_{\rm FIR})= 8.997+1.408\times(\log(L_{\rm IR})-10)$)
derived for local galaxies in \citet{lut16}.  The derived physical
sizes and surface IR luminosities are summarized in
Table~\ref{tbl:prop}.

\subsection{Progenitors of $z\gtrsim3$ compact quiescent galaxies?}

Based on their mm-wave sizes and redshift estimates, \citet{ika14}
suggested that ASXDF1100.053.1 and 231.1 may be the progenitors of
$z\sim2$ compact quiescent galaxies (cQGs), the evolutionary scenario
suggested by \citet{tof14}.  cQGs have now reported out to $z\sim4$
\citep{str15}.  Their stellar components have
$R_{\rm c,e}= 0.3$--3.2\,kpc, with the median of 0.63$\pm$0.18\,kpc.  ASXDF1100.053.1 and 231.1 have compact enough
starburst regions to evolve into cQGs at $z\sim3$--4.  ASXDF1100.053.1
has already created a stellar mass comparable to that found in cQGs at
$z\sim2$, $[0.4-5]\times10^{11}$\,${\rm M_{\odot}}$
\citep[e.g.][]{bell14,kro14}, and at $z\sim3$--4,
$[0.4-1.8]\times10^{11}$\,${\rm M_{\odot}}$ \citep{str15}, based on a
Chabrier IMF \citep{chab03}.  Furthermore, given the SMG duty cycle of
$t_{\rm burst}=42^{+40}_{-29}$\,Myr suggested in \citet{tof14}, and
the derived SFRs by mm measurements, its stellar mass is
expected to increase by $\approx[0.8-6]\times10^{10}$\,$M_{\odot}$.

The observed stellar mass of ASXDF1100.231.1 is small in comparison
with known cQGs, although there are large uncertainties.
ASXDF1100.231.1 will generate an additional
$\approx[0.5-4]\times10^{10}$\,$M_{\odot}$ of stars via its on-going
star formation, and can thus become similarly massive to the known
cQGs.

From these facts, it seems clear that ASXDF1100.053.1 and 231.1 can
potentially evolve into cQGs at $z\gtrsim3$, although it depends on
their remaining gas masses and whether they quench their star
formations on short timescales, i.e.\ the $t_{\rm burst}$ noted above.
 
\section{Summary}

We have conducted a detailed multi-wavelength study of two
ALMA-identified SMGs, ASXDF1100.053.1 and 231.1, with extremely red
submm colors, aiming to constrain their redshifts and better
understand their nature.  Based on their radio/submm colors,
we have determined redshifts of $6.5^{+1.4}_{-1.1}$ and
$4.1^{+0.6}_{-0.7}$ for ASXDF1100.053.1 and 231.1, respectively.  We
have quantified the influence of the CMB on these photometric
redshifts using simple models, finding that even at $z\gtrsim6$ the
effects of the CMB do not lead to significant over-estimation.

We have measured mm-wave sizes of ASXDF1100.053.1 and 231.1 in deep
ALMA continuum images.  The derived circularized half-light radii of
ASXDF1100.053.1 and 231.1 are $\sim1$ and $\lesssim0.2$\,kpc,
respectively.  Their surface IR luminosity densities are
$\sim1\times10^{12}$ and
$\gtrsim1.5\times10^{13}$\,$L_{\odot}$\,kpc$^{-2}$, comparable to
those of local (U)LIRGs and consistent with a known empirical trend in
$L_{\rm IR}-\Sigma_{\rm L_{IR}}$ seen for local galaxies.

From an optical/near-/mid-IR SED analysis of ASXDF1100.053.1
and 231.1, adopting their radio/submm photometric redshifts.
we find that ASXDF1100.231.1 has near-/mid-IR colors
consistent with the existence of a redshifted H$\alpha$ line at
$z=4$--5 in the IRAC 3.6-$\mu$m band.  The derived stellar masses of
ASXDF1100.053.1 and 231.1 are comparable to those of known SMGs.

Given the observed stellar masses, SFRs and typical cycle times of
SMGs, we find that ASXDF1100.053.1 and 231.1 can evolve into cQGs at
$z\gtrsim3$.  Our intensive studies of SMGs at $z\gtrsim4$, using the
new capabilities of ALMA and JVLA, have allowed us to discover
plausible candidate $z\gtrsim6$ and $z\sim4$ SMGs which are too
heavily dust-obscured to be detected even in the deepest
optical/near-/mid-IR images.

\begin{table*}
\begin{center}
\caption{Derived properties of ASXDF1100.053.1 and 231.1.\label{tbl:prop}}
\scalebox{1.0}{ 
\begin{tabular}{ c c c c c  c c c c}
\tableline \tableline 
              &       stellar mass                                                           & $\chi^2_{\rm best}$  &    \multicolumn{2}{c}{$L_{\rm IR}$(8--1000\,$\mu$m) } &    \multicolumn{2}{c}{SFR(IR) } 	                                  &  $R_{\rm c,e}$                     &  $\Sigma_{L_{\rm IR}}$\\  \cline{4-5} \cline{6-7}
              &           						                                         &				      &       w/o CMB 	              &	  w/ CMB   		        &	 w/o CMB                        &    w/ CMB                          &                                             &  \\
              &                                        ${\rm M_{\odot}}$            &                                  &    10$^{12}$\,${\rm L_{\odot}}$ &     10$^{12}$\,${\rm L_{\odot}}$  &    ${\rm M_{\odot}}$\,yr$^{-1}$  &     ${\rm M_{\odot}}$\,yr$^{-1}$      &   kpc                                 & ${\rm L_{\odot}}$\,kpc$^{-1}$ \\ \tableline
                                                     \multicolumn{9}{c}{ASXDF1100.053.1 w/o emission lines } \\  \tableline 
 $z=5.5$   &     $1.2^{+2.5}_{-0.81}$\,$\times10^{11}$       &   3.4    		       &  5.8\,$\pm0.25$          &      6.5\,$\pm0.28$                &     580\,$\pm25$         &    650\,$\pm28$    	          &	$1.05^{+0.06}_{-0.03}$   &   $1.0^{+0.05}_{-0.07}$\,$\times10^{12}$   \\
 $z=6.5$   &     $1.0^{+4.8}_{-0.90}$\,$\times10^{11}$   &   4.5     	               &  5.8\,$\pm0.25$             &    6.8\,$\pm0.29$              &       580\,$\pm25$       &     680\,$\pm29$   	          &	$0.95^{+0.05}_{-0.03}$  &   $1.2^{+0.06}_{-0.07}$\,$\times10^{12}$   \\  
 $z=7.5$   &    $1.6^{+10}_{-1.4}$\,$\times10^{11}$    &      4.4     		       &    6.0\,$\pm0.26$              &    7.4\,$\pm0.32$                  &       600\,$\pm26$         &    740\,$\pm32$   	          &	$0.88^{+0.05}_{-0.03}$   &   $1.5^{+0.08}_{-0.09}$\,$\times10^{12}$ \\                 
 \tableline 
                                                     \multicolumn{9}{c}{ASXDF1100.053.1 w/  emission lines } \\  \tableline 
 $z=5.5$   & $1.0^{+2.4}_{-0.69}$\,$\times10^{11}$          &     2.8                     &  5.8\,$\pm0.25$                &      6.5\,$\pm0.28$                 &      580\,$\pm25$          &    650\,$\pm28$     	          &	$1.05^{+0.06}_{-0.03}$   &    $1.0^{+0.05}_{-0.07}$\,$\times10^{12}$   \\
 $z=6.5$   &     $8.6^{+42}_{-7.3}$\,$\times10^{10}$   &    4.2     	               &   5.8\,$\pm0.25$               &    6.8\,$\pm0.29$               &       580\,$\pm25$         &      680\,$\pm29$  	           &	$0.95^{+0.05}_{-0.03}$   &    $1.2^{+0.06}_{-0.07}$\,$\times10^{12}$    \\  
 $z=7.5$   &    $1.7^{+11}_{-1.6}$\,$\times10^{11}$    &     3.8       		       &    6.0\,$\pm0.26$               &    7.4\,$\pm0.32$                 &       600\,$\pm26$          &     740\,$\pm32$     	           &	$0.88^{+0.05}_{-0.03}$  &   $1.5^{+0.08}_{-0.09}$\,$\times10^{12}$    \\                 
 \tableline 
                                                      \multicolumn{9}{c}{ASXDF1100.231.1 w/o emission lines } \\  \tableline 
  $z=3.5$   &     $4.5^{+8.1}_{-3.9}$\,$\times10^{9}$       &      12    		       &  4.3\,$\pm0.15$                 &      4.5\,$\pm0.16$                &    430\,$\pm15$         &    450\,$\pm16$     	            &	$\leq0.25$                       &    $\geq$1.5$\times10^{13}$  \\
  $z=4.5$   &     $2.6^{+3.1}_{-1.5}$\,$\times10^{10}$   &      10   		       &  3.9\,$\pm0.14$                &   4.2\,$\pm0.15$                &       390\,$\pm14$          &     420\,$\pm15$  		    &	$\leq0.22$                       &    $\geq$1.7$\times10^{13}$    \\  
  $z=5.5$   &    $3.3^{+3.6}_{-1.9}$\,$\times10^{10}$    &       7.5        	       &    3.8\,$\pm0.13$              &    4.2\,$\pm0.15$                 &       380$\pm13$         &     420\,$\pm15$    	 	    &	$\leq0.20$                       &     $\geq$2.1$\times10^{13}$    \\                 
 \tableline 
                                                     \multicolumn{9}{c}{ASXDF1100.231 w/  emission lines } \\  \tableline 
  $z=3.5$   & $4.3^{+7.8}_{-3.7}$\,$\times10^{9}$          &       12    	                &  4.3\,$\pm0.15$                  &      4.5\,$\pm0.16$                  &      430\,$\pm15$           &    450\,$\pm16$     	            &	$\leq0.25$                       &   $\geq$1.5$\times10^{13}$     \\
  $z=4.5$   &     $2.6^{+2.3}_{-1.2}$\,$\times10^{10}$    &     5.5    		        &   3.9\,$\pm0.14$                 &    4.2\,$\pm0.15$                   &       390\,$\pm14$          &     420\,$\pm15$    		    &	$\leq0.22$                       &  $\geq$1.7$\times10^{13}$        \\  
  $z=5.5$   &    $2.8^{+3.3}_{-1.6}$\,$\times10^{10}$     &     8.4      		&    3.8\,$\pm0.13$               &    4.2\,$\pm0.15$                 &       380$\pm13$          &     420\,$\pm15$    	             &	$\leq0.20$               &   $\geq$2.1$\times10^{13}$       \\      
    \tableline           
\end{tabular}}
 \end{center}
\end{table*}

\acknowledgments This paper makes use of the following ALMA data:
ADS/JAO.ALMA\#2012.1.00326.S. and 2013.1.00781.S. ALMA is a
partnership of ESO (representing its member states), NSF (USA)
and NINS (Japan), together with NRC (Canada) and NSC and ASIAA
(Taiwan), in cooperation with the Republic of Chile. The Joint
ALMA Observatory is operated by ESO, AUI/NRAO and NAOJ. SI and
KIC acknowledge the support of the Netherlands Organization for
Scientific Research (NWO) through the Top Grant Project
614.001.403. RJI acknowledges support from ERC in the form of the
Advanced Grant, 321302, COSMICISM. YT is supported by JSPS
KAKENHI (No.\ 15H02073). 

{\it Facilities:} \facility{ALMA}, \facility{VLA}, \facility{ASTE}, \facility{Herschel},\facility{Spitzer},\facility{UKIRT},\facility{Subaru}

\appendix

\section{Simulations of enclosed fluxes with Gaussian models}\label{sec:appendix}
Here we describe simple sanity checks of our measurements of enclosed fluxes in the ALMA clean component maps presented in \S~\ref{sec:size}. 
For ASXDF1100.053.1. 
We prepared 100 independent noise data of ALMA visibility generated from actual ALMA data for ASXDF1100.053.1. 
We input a symmetric Gaussian model with a size into the 100 noise visibility data. 
Here a total flux of the input model is same with that of ASXDF1100.053.1. 
We imaged and cleaned the 100 noise visibilities with the input model in the same manner with the actual observed data for ASXDF1100.053.1, and we measured enclosed fluxes. 
We repeated these process from input sizes of 0$''$.025 (FWHM) to 0$''$.800 with a step of 0$''$.025.

 \begin{figure*}
\epsscale{1.1}
\plotone{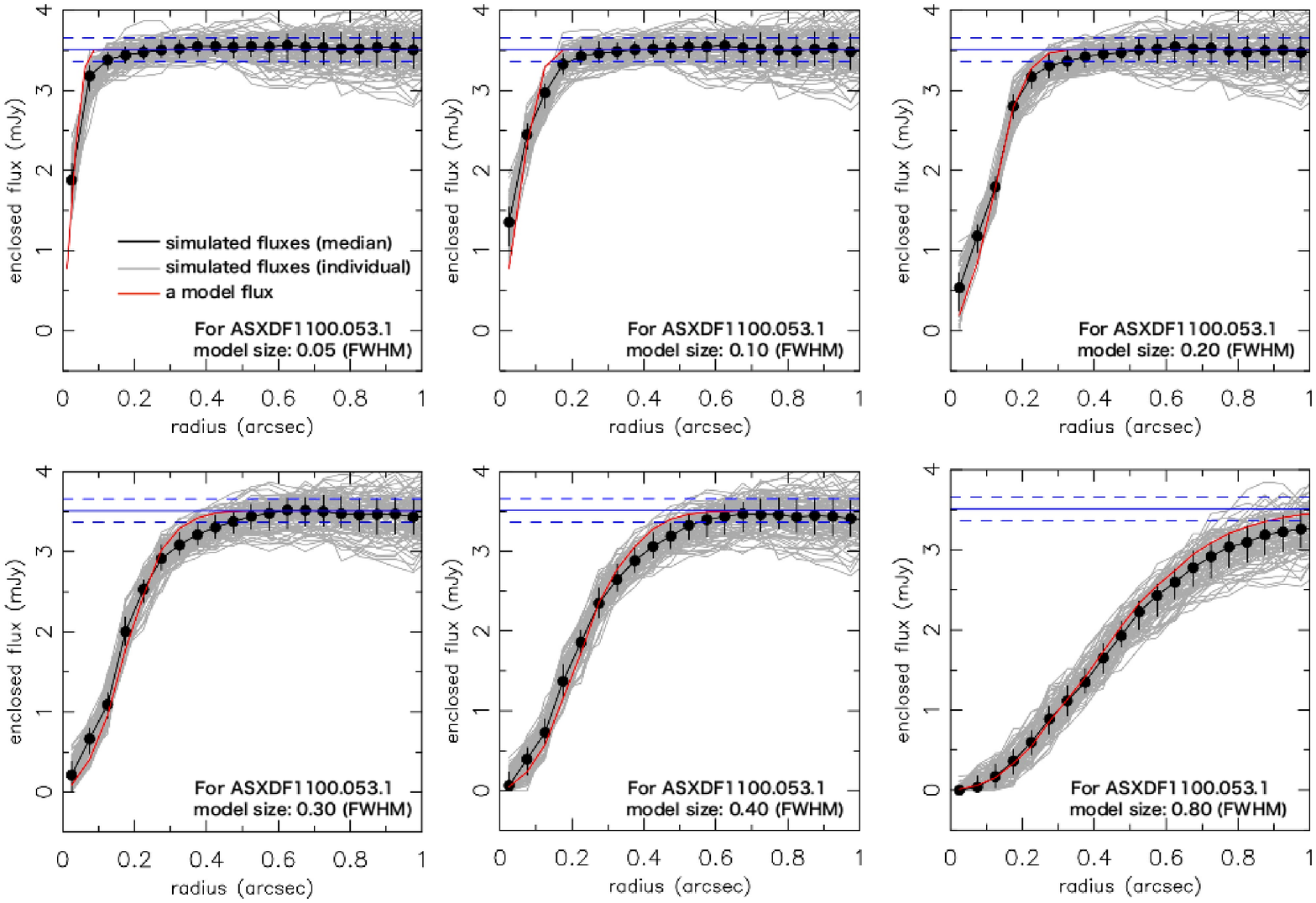}
\caption{
Simulated enclosed fluxes using symmetric Gaussian models as a function of radius for ALMA data of ASXDF1100.053.1. 
Black dots and vertical lines mark medians and 1$\sigma$ dispersions of the 100 simulated fluxes. 
Grey lines show enclosed fluxes of the 100 mock sources. 
A red line marks a raw flux of an input Gaussian model with a spatial sampling of 0$''$.05\,pixel$^{-1}$ (for a model size of 0$''$.05, a sampling of 0$''$.025\,pixel$^{-1}$).  
Blue solid and dashed lines mark the total flux and error of ASXDF1100.053.1 measured in the ALMA beam-convolved image independently. 
Here we plot for model sizes of 0$''$.05, 0$''$.10, 0$''$.20, 0$''$.30, 0$''$.40, and 0$''$.80 (FWHM) as representatives among all simulated sizes.}   \label{fig:simu053}
  \vspace{-0mm}
\end{figure*}

Fig.~\ref{fig:simu053} shows resulted enclosed fluxes of mock Gaussian sources as a function of radius. 
The extracted fluxes of the mock sources show that the input total fluxes are recovered. 
However, in the cases of mock sources with larger input-size, the input total fluxes are not completely recovered. 
A fraction of flux remaining in a residual image is expected to get larger in our clean process down to the $1\sigma$ noise level for a source with a larger size. 
At a mm-wave size of $\lesssim0''$.40 where ASXDF1100.053.1 is located, an expected missing flux is $\lesssim$2\%, which is negligible. 
We also conducted simulations using the ALMA data for ASXDF1100.231.1 and we obtained similar results to those for ASXDF1100.053.1.

\begin{figure}
   \vspace{-0mm}
\epsscale{1.1}
\plotone{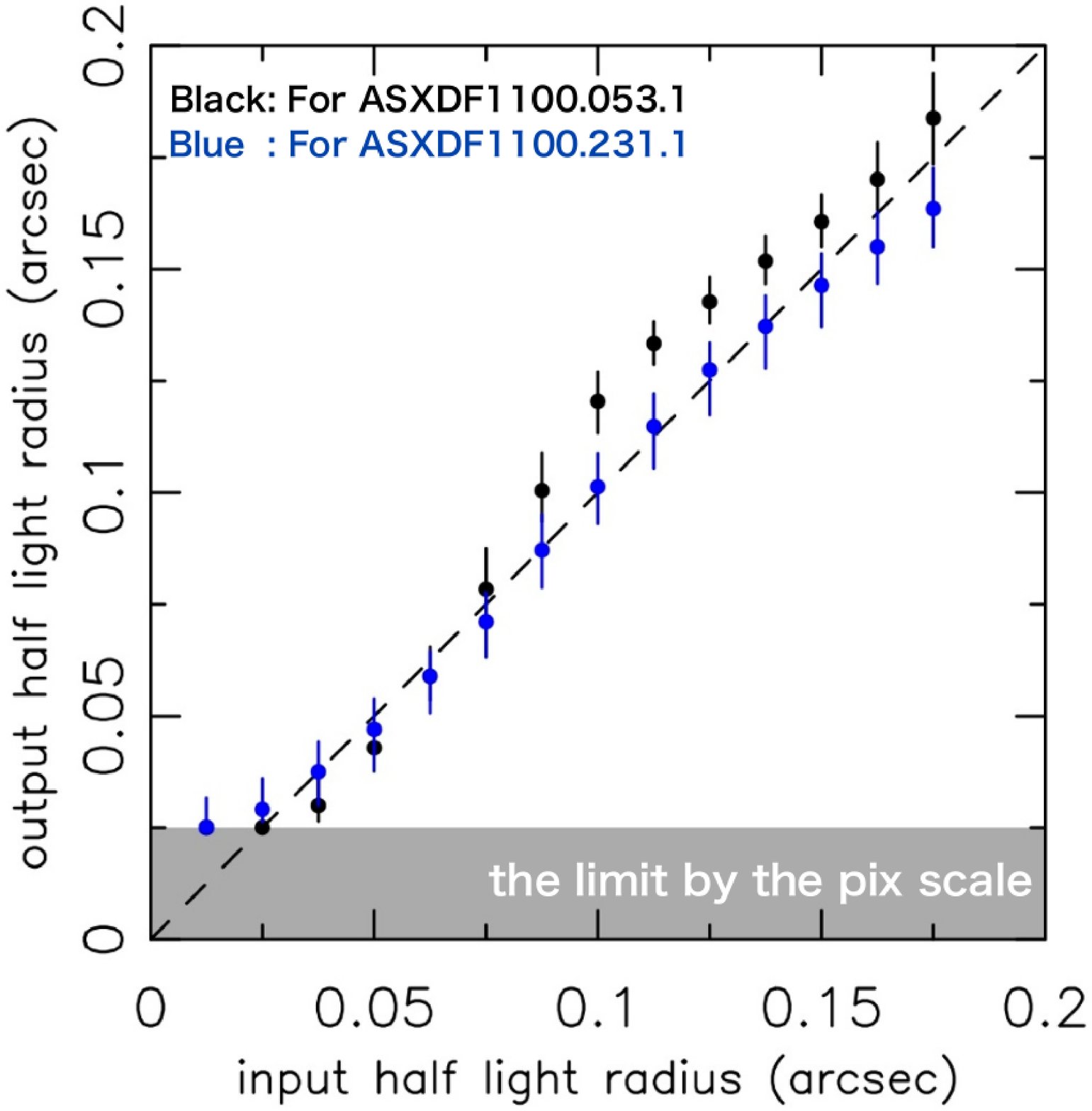}
\caption{
Comparisons of input and output half light radii derived from Montecarlo simulations using symmetric Gaussian models. 
These simulations used the noise visibility data which were generated from the actual ALMA data for ASXDF1100.053.1 (Black) and 231.1 (Blue). 
Output size, measured half light  radii were obtained with the same manner with real measurements. 
A grey shaded region marks the limit for output sizes due to a pixel scale of 0$''$.05\,pixel$^{-1}$. 
}   \label{fig:compsize053}
\end{figure}

Fig.~\ref{fig:compsize053} shows comparisons between input and output half light radii from the simulations based on the ALMA data for ASXDF1100.053.1 and 231.1, respectively. 
Both plots indicate that our source size measurements in the ALMA continuum clean component maps are sensitive down to the limit by the pixel scale of 0$''$.05\,pixel$^{-1}$. 
Fig.~\ref{fig:compsize053} indicates that source size measurements in the ALMA continuum clean component map of ASXDF1100.231.1 can reproduce sizes well. 
On the other hand, Fig.~\ref{fig:compsize053} shows that size measurements in the ALMA continuum clean component map of ASXDF1100.053.1 can overestimate sizes around a size of $\sim0.$10--0.15 arcsec. 
The overestimation of sizes can be at most $\sim$0.02 arcsec or 20\%. 
However, the measured size ($R_{\rm c,e}$) of ASXDF1100.053 is 0$''$.17, therefore,  the possible systematic offset is expected not to contribute to the measured size of ASXDF1100.053.1.

\begin{thebibliography}{}
\bibitem[Aretxaga et al.(2003)]{are03} Aretxaga, I., Hughes, D.~H., Chapin, E.~L., et al.\ 2003, \mnras, 342, 759 
\bibitem[Aretxaga et al.(2005)]{are05} Aretxaga, I., Hughes, D.~H., \& Dunlop, J.~S.\ 2005, \mnras, 358, 1240 
\bibitem[Aretxaga et al.(2007)]{are07} Aretxaga, I., Hughes, D.~H., Coppin, K., et al.\ 2007, \mnras, 379, 1571
\bibitem[Aretxaga et al.(2011)]{are11} Aretxaga, I., Wilson, G.~W., Aguilar, E., et al.\ 2011, \mnras, 415, 3831 
\bibitem[Arnouts et al.(1999)]{arn99} Arnouts, S., Cristiani, S., Moscardini, L., et al.\ 1999, \mnras, 310, 540 
\bibitem[Arumugam et al.(2016)]{aru16} Arumugam, V., et al.\ 2016, submitted  
\bibitem[Asboth et al.(2016)]{asb16} Asboth, V., Conley, A., Sayers, J., et al.\ 2016, \mnras, 462, 1989 
\bibitem[Ashby et al.(2013)]{ash13} Ashby, M.~L.~N., Willner, S.~P., Fazio, G.~G., et al.\ 2013, \apj, 769, 80 
\bibitem[Barro et al.(2016)]{bar16} Barro, G., Kriek, M., P{\'e}rez-Gonz{\'a}lez, P.~G., et al.\ 2016, \apjl, 827, L32 
\bibitem[Bell(2003)]{bel03} Bell, E.~F.\ 2003, \apj, 586, 794 
\bibitem[Belli et al.(2014)]{bell14} Belli, S., Newman, A.~B., Ellis, R.~S., \& Konidaris, N.~P.\ 2014, \apjl, 788, L29 
\bibitem[Biggs et al.(2011)]{big11} Biggs, A.~D., Ivison, R.~J., Ibar, E., et al.\ 2011, \mnras, 413, 2314 
\bibitem[Blain et al.(2002)]{bla02} Blain, A.~W., Smail, I., Ivison, R.~J., Kneib, J.-P., \& Frayer, D.~T.\ 2002, \physrep, 369, 111
\bibitem[Borys et al.(2004)]{bor04} Borys, C., Scott, D., Chapman, S., et al.\ 2004, \mnras, 355, 485  
\bibitem[Bruzual \& Charlot(2003)]{bru03} Bruzual, G., \& Charlot, S.\ 2003, \mnras, 344, 1000 
\bibitem[Calzetti et al.(2000)]{cal00} Calzetti, D., Armus, L., Bohlin, R.~C., et al.\ 2000, \apj, 533, 682 
\bibitem[Capak et al.(2011)]{capa11} Capak, P.~L., Riechers, D., Scoville, N.~Z., et al.\ 2011, \nat, 470, 233 
\bibitem[Caputi et al.(2011)]{cap11} Caputi, K.~I., Cirasuolo, M., Dunlop, J.~S., et al.\ 2011, \mnras, 413, 162
\bibitem[Caputi et al.(2014)]{cap14} Caputi, K.~I., Michalowski, M., Krips, M., et al. \ 2014, \apj, 788, 126
\bibitem[Carilli \& Yun(1999)]{car99} Carilli, C.~L., \& Yun, M.~S.\ 1999, \apjl, 513, L13
\bibitem[Carilli et al.(2011)]{car11} Carilli, C.~L., Hodge, J., Walter, F., et al.\ 2011, \apjl, 739, L33 
\bibitem[Chabrier(2003)]{chab03} Chabrier, G.\ 2003, \pasp, 115, 763 
\bibitem[Chapman et al.(2003)]{chap03} Chapman, S.~C., Blain, A.~W., Ivison, R.~J., \& Smail, I.~R.\ 2003, \nat, 422, 695
\bibitem[Chapman et al.(2005)]{cha05} Chapman, S.~C., Blain, A.~W., Smail, I., \& Ivison, R.~J.\ 2005, \apj, 622, 772 
\bibitem[Chary \& Elbaz(2001)]{cha03} Chary, R., \& Elbaz, D.\ 2001, \apj, 556, 562 
\bibitem[Condon(1992)]{con92} Condon, J.~J.\ 1992, \araa, 30, 575
\bibitem[Cowley et al.(2015)]{wil15} Cowley, W.~I., Lacey, C.~G., Baugh, C.~M., \& Cole, S.\ 2015, \mnras, 446, 1784 
\bibitem[Cox et al.(2011)]{cox11} Cox, P., Krips, M., Neri, R., et al.\ 2011, \apj, 740, 63 
\bibitem[da Cunha et al.(2013)]{dac13} da Cunha, E., Groves, B., Walter, F., et al.\ 2013, \apj, 766, 13 
\bibitem[da Cunha et al.(2015)]{dac15} da Cunha, E., Walter, F., Smail, I.~R., et al.\ 2015, \apj, 806, 110 
\bibitem[Dannerbauer et al.(2008)]{dan08} Dannerbauer, H., Walter, F., \& Morrison, G.\ 2008, \apjl, 673, L127 
\bibitem[Dowell et al.(2014)]{dow14} Dowell, C.~D., Conley, A., Glenn, J., et al.\ 2014, \apj, 780, 75 
\bibitem[Draine et al.(2007)]{dra07} Draine, B.~T., Dale, D.~A., Bendo, G., et al.\ 2007, \apj, 663, 866 
\bibitem[Furusawa et al.(2008)]{fur08} Furusawa, H., Kosugi, G., Akiyama, M., et al.\ 2008, \apjs, 176, 1 
\bibitem[Geach et al.(2016)]{gea16} Geach, J.~E., Dunlop, J.~S., Halpern, M., et al.\ 2016, arXiv:1607.03904 
\bibitem[Gear et al.(2000)]{gea00} Gear, W.~K., Lilly, S.~J., Stevens, J.~A., et al.\ 2000, \mnras, 316, L51  
\bibitem[Harrison et al.(2016)]{har16} Harrison, C.~M., Simpson, J.~M., Stanley, F., et al.\ 2016, \mnras, 457, L122 
\bibitem[Hayward et al.(2013)]{hay13} Hayward, C.~C., Narayanan, D., Kere{\v s}, D., et al.\ 2013, \mnras, 428, 2529 
\bibitem[Hodge et al.(2013)]{hod13} Hodge, J.~A., Karim, A., Smail, I., et al.\ 2013, \apj, 768, 91
\bibitem[Hodge et al.(2016)]{hod16} Hodge, J.~A., Swinbank, A.~M., Simpson, J.~M., et al.\ 2016, \apj, 833, 103   
\bibitem[Hughes et al.(1998)]{hug98} Hughes, D.~H., Serjeant, S., Dunlop, J., et al.\ 1998, \nat, 394, 241 
\bibitem[Hughes et al.(2002)]{hug02} Hughes, D.~H., Aretxaga, I., Chapin, E.~L., et al.\ 2002, \mnras, 335, 871 
\bibitem[Ikarashi et al.(2011)]{ika11} Ikarashi, S., Kohno, K., Aguirre, J.~E., et al.\ 2011, \mnras, 415, 3081 
\bibitem[Ikarashi et al.(2015)]{ika14} Ikarashi, S., Ivison, R.~J., Caputi, K.~I., et al.\ 2015, \apj, 810, 133 
\bibitem[Ilbert et al.(2006)]{ilb06} Ilbert, O., Arnouts, S., McCracken, H.~J., et al.\ 2006, \aap, 457, 841 
\bibitem[Iono et al.(2006)]{ion06} Iono, D., Peck, A.~B., Pope, A., et al.\ 2006, \apjl, 640, L1 
\bibitem[Iono et al.(2016)]{ion16} Iono, D., Yun, M.~S., Aretxaga, I., et al.\ 2016, \apjl, 829, L10 
\bibitem[Isobe et al.(1986)]{iso86} Isobe, T., Feigelson, E.~D., \& Nelson, P.~I.\ 1986, \apj, 306, 490 
\bibitem[Ivison et al.(1998)]{ivi98} Ivison, R.~J., Smail, I., Le Borgne, J.-F., et al.\ 1998, \mnras, 298, 583 
\bibitem[Ivison et al.(2000)]{ivi00} Ivison, R.~J., Smail, I., Barger, A.~J., et al.\ 2000, \mnras, 315, 209 
\bibitem[Ivison et al.(2002)]{ivi02} Ivison, R.~J., Greve, T.~R., Smail, I., et al.\ 2002, \mnras, 337, 1 
\bibitem[Ivison et al.(2005)]{ivi05} Ivison, R.~J., Smail, I., Dunlop, J.~S., et al.\ 2005, \mnras, 364, 1025
\bibitem[Ivison et al.(2007)]{ivi07} Ivison, R.~J., Greve, T.~R., Dunlop, J.~S., et al.\ 2007, \mnras, 380, 199
\bibitem[Ivison et al.(2010)]{ivi10} Ivison, R.~J., Swinbank, A.~M., Swinyard, B., et al.\ 2010, \aap, 518, L35 
\bibitem[Ivison et al.(2016)]{ivi16} Ivison, R.~J., Lewis, A.~J.~R., Weiss, A., et al.\ 2016, \apj, 832, 78 
\bibitem[Kennicutt(1998)]{ken98} Kennicutt, R.~C., Jr.\ 1998, \araa, 36, 189   
\bibitem[Krogager et al.(2014)]{kro14} Krogager, J.-K., Zirm, A.~W., Toft, S., Man, A., \& Brammer, G.\ 2014, \apj, 797, 17
\bibitem[Lawrence et al.(2007)]{law07} Lawrence, A., Warren, S.~J., Almaini, O., et al.\ 2007, \mnras, 379, 1599
\bibitem[Lindner et al.(2011)]{lin11} Lindner, R.~R., Baker, A.~J., Omont, A., et al.\ 2011, \apj, 737, 83 
\bibitem[Lutz et al.(2016)]{lut16} Lutz, D., Berta, S., Contursi, A., et al.\ 2016, \aap, 591, A136 
\bibitem[McBride et al.(2014)]{mcb14} McBride, J., Quataert, E., Heiles, C., \& Bauermeister, A.\ 2014, \apj, 780, 182 
\bibitem[Messias et al.(2014)]{mes14} Messias, H., Dye, S., Nagar, N., et al.\ 2014, \aap, 568, A92
\bibitem[Murphy(2009)]{mur09} Murphy, E.~J.\ 2009, \apj, 706, 482
\bibitem[Oliver et al.(2012)]{oli12} Oliver, S.~J., Bock, J., Altieri, B., et al.\ 2012, \mnras, 424, 1614 
\bibitem[Oteo et al.(2016)]{ote16} Oteo, I., Zwaan, M.~A., Ivison, R.~J., Smail, I., \& Biggs, A.~D.\ 2016, arXiv:1607.06464 
\bibitem[Pope et al.(2006)]{pop06} Pope, A., Scott, D., Dickinson, M., et al.\ 2006, \mnras, 370, 1185 
\bibitem[Riechers et al.(2013)]{rie13} Riechers, D.~A., Bradford, C.~M., Clements, D.~L., et al.\ 2013, \nat, 496, 329 
\bibitem[Rieke et al.(2009)]{rie09} Rieke, G.~H., Alonso-Herrero, A., Weiner, B.~J., et al.\ 2009, \apj, 692, 556 
\bibitem[Silva et al.(1998)]{sil98} Silva, L., Granato, G.~L., Bressan, A., \& Danese, L.\ 1998, \apj, 509, 103 
\bibitem[Simpson et al.(2014)]{sim14} Simpson, J.~M., Swinbank, A.~M., Smail, I., et al.\ 2014, \apj, 788, 125
\bibitem[Simpson et al.(2015a)]{sim15a} Simpson, J.~M., Smail, I., Swinbank, A.~M., et al.\ 2015a, \apj, 799, 81 
\bibitem[Simpson et al.(2015b)]{sim15b} Simpson, J.~M., Smail, I., Swinbank, A.~M., et al.\ 2015b, \apj, 807, 128
\bibitem[Smail et al.(1997)]{sib97} Smail, I., Ivison, R.~J., \& Blain, A.~W.\ 1997, \apjl, 490, L5 
\bibitem[Smail et al.(1999)]{sma99} Smail, I., Ivison, R.~J., Kneib, J.-P., et al.\ 1999, \mnras, 308, 1061  
\bibitem[Smol{\v c}i{\'c} et al.(2012)]{smo12} Smol{\v c}i{\'c}, V., Aravena, M., Navarrete, F., et al.\ 2012, \aap, 548, A4 
\bibitem[Smol{\v c}i{\'c} et al.(2015)]{smo15} Smol{\v c}i{\'c}, V., Karim, A., Miettinen, O., et al.\ 2015, \aap, 576, A127 
\bibitem[Straatman et al.(2015)]{str15} Straatman, C.~M.~S., Labb{\'e}, I., Spitler, L.~R., et al.\ 2015, \apjl, 808, L29 
\bibitem[Strandet et al.(2016)]{str16} Strandet, M.~L., Weiss, A., Vieira, J.~D., et al.\ 2016, \apj, 822, 80 
\bibitem[Swinbank et al.(2012)]{swi12} Swinbank, A.~M., Karim, A., Smail, I., et al.\ 2012, \mnras, 427, 1066 
\bibitem[Swinbank et al.(2014)]{swi14} Swinbank, A.~M., Simpson, J.~M., Smail, I., et al.\ 2014, \mnras, 438, 1267 
\bibitem[Toft et al.(2014)]{tof14} Toft, S., Smol{\v c}i{\'c}, V., Magnelli, B., et al.\ 2014, \apj, 782, 68 
\bibitem[Vieira et al.(2013)]{vie13} Vieira, J.~D., Marrone, D.~P., Chapman, S.~C., et al.\ 2013, \nat, 495, 344
\bibitem[Walter et al.(2012)]{wal12} Walter, F., Decarli, R., Carilli, C., et al.\ 2012, \nat, 486, 233
\bibitem[Wang et al.(2009)]{wan09} Wang, W.-H., Barger, A.~J., \& Cowie, L.~L.\ 2009, \apj, 690, 319 
\bibitem[Wardlow et al.(2011)]{war11} Wardlow, J.~L., Smail, I., Coppin, K.~E.~K., et al.\ 2011, \mnras, 415, 1479  
\bibitem[Wardlow et al.(2013)]{war13} Wardlow, J.~L., Cooray, A., De Bernardis, F., et al.\ 2013, \apj, 762, 59
\bibitem[Wei{\ss} et al.(2009)]{wei09a} Wei{\ss}, A., Ivison, R.~J., Downes, D., et al.\ 2009, \apjl, 705, L45
\bibitem[Wei{\ss} et al.(2013)]{wei13} Wei{\ss}, A., De Breuck, C., Marrone, D.~P., et al.\ 2013, \apj, 767, 88 
\bibitem[Younger et al.(2007)]{you07} Younger, J.~D., Fazio, G.~G., Huang, J.-S., et al.\ 2007, \apj, 671, 1531  
\bibitem[Younger et al.(2009)]{you09} Younger, J.~D., Fazio, G.~G., Huang, J.-S., et al.\ 2009, \apj, 704, 803 
\bibitem[Yun et al.(2001)]{yun01} Yun, M.~S., Reddy, N.~A., \& Condon, J.~J.\ 2001, \apj, 554, 803 
\bibitem[Yun et al.(2012)]{yun12} Yun, M.~S., Scott, K.~S., Guo, Y., et al.\ 2012, \mnras, 420, 957
\bibitem[Zhang et al.(2016)]{zha16} Zhang, Z.-Y., Papadopoulos, P.~P., Ivison, R.~J., et al.\ 2016, Royal Society Open Science, 3, 160025
\end{thebibliography}
\end{document}